\documentclass{aa}  
\usepackage{graphicx}
\usepackage{hyperref}
\usepackage{txfonts}
\begin{document}

   \title{The polarization-angle flip in GRB prompt emission}

   \author{Kangfa Cheng
          \inst{1}\fnmsep 
          \thanks{}
          \and
          Xiaohong Zhao \inst{2,3,4}
          \and 
          Jirong Mao
          \inst{2,3,4}
          \and
          Zhifu Chen \inst{1}
          }

   \institute{ School of Mathematics and Physics, Guangxi Minzu University, Nanning 530006, China \\
          \email{chengkf@gxmzu.edu.cn}
         \and
         Yunnan Observatories, Chinese Academy of Sciences, Kunming, 650011, People’s Republic of China \\
         \and
          Key Laboratory for the Structure and Evolution of Celestial Objects, Chinese Academy of Science, People’s Republic of China \\
         \and 
          Center for Astronomical Mega-Science, Chinese Academy of Science, 20A Datun Road, Chaoyang District, Beijing 100012, People’s Republic of China
             }

   \date{Received XXX; accepted XXX}

 
  \abstract
   {In recent years, some polarization measurements of gamma-ray bursts (GRBs) have been reported, and the polarization-angle (PA) rotation in the prompt emission phase has been found in several bursts. The physical mechanism of the PA evolution is still unclear. In this work, we studied the origin of the PA rotation in a toroidal magnetic field.} 
  {We aim to provide an explanation for the PA rotation in GRBs and find the physical conditions that lead to the rotation by 90 degrees in the toroidal magnetic-field (MF) model. Moreover, we present some observable polarization properties in the MF model that can be tested in the future.
}
   {We calculated the instantaneous polarization degree (PD) from a top-hat jet with different normalized viewing angles ($q=\theta_v/ \theta_j$), jet opening angles ($\theta_j$), and jet Lorentz factors ($\Gamma$) in three wavebands. When the PD changes between positive and negative values, it means that the PA flips by 90 degrees. On these grounds, we can summarize the range of parameters required for these PA flips. Considering these parameter conditions, we can further estimate the observed rate of the GRBs exhibiting such PA rotations.
   }
     {We find that the PA rotation in the toroidal MF is primarily related to three critical factors: the viewing angle, the jet opening angle, and the jet Lorentz factor. Additionally, the PA can experience flips of 90 degrees twice. The conditions for the flips are $q \gtrsim 0.5$ (except for $q\simeq 1$) and $y_j =(\Gamma \theta_j)^2 \gtrsim 4$. However, the two flips in the PA might not be concurrently observable due to the constraint of flux. Taking these conditions into account and assuming a random orientation between the jet axis and the line of sight (LOS), we obtain a theoretical upper limit (without any constraints) for the observed rate of GRBs in the X-ray or $\gamma$-ray band displaying the flips in PA as $R_{ch} \lesssim 80\%$. We further constrain the observed rate as $R_{ch} \sim 16\%$ according to the maximal post-flip polarized flux level, where the observed rate of single and double flips each account for $\sim 8\%$. It should be noted that the observed rates are different in various wavebands. The observed rate of the second PA flip in the optical bands should be higher than that in the X-ray or $\gamma$-ray band since the flux in the optical band declines much slower than that in the X-ray or $\gamma$-ray band. Moreover, when the LOS is close to the jet edge ($q\to 1$), it is the easiest case in which to observe the 90-degree PA flip due to the relatively high post-flip polarized flux level. The first and second PA flips in a GRB pulse are most likely to occur at the observed times of $t_{obs} \sim [2-3] t_{peak}$ and $\sim [3-4] t_{peak}$, respectively, where $t_{peak}$ is the peak time of the pulse. It is also noted that the PA flip would not happen before the peak time.}
     

   \maketitle
%

\section{Introduction}
\label{sec:1}

Polarization measurements play a crucial role in shedding light on the composition of gamma-ray-burst (GRB) ejecta, jet structure, and the viewing geometry of a GRB jet. They are also vital for understanding the radiation mechanism and magnetic field (MF) structure in GRB ejecta.  There are two possible MF models including the large-scale ordered MF and small-scale random MF, which have different origins. The small-scale random MF generally originates from Weibel instability in the shock \citep{Gruzinov+Waxman+1999, Medvedev+Loeb+1999} or turbulence \citep{mao11,Mao+Wang+2013}, while the large-scale ordered MF originates from the central objects of GRB \citep{Spruit+2001, Lazzati+2006}. The two MF models will lead to quite different polarization properties in observations if the radiation mechanism is synchrotron emission. The polarization degree (PD) in the large-scale ordered MF case would be much higher than that in the small-scale random MF case, and the PD evolution is different in the two cases. Thus, polarization can be used as a probe to investigate the MF configuration of GRBs \citep{Granot+konigl+2003, Nakar+etal+2003, Toma+etal+2009, Mao+Wang+2013, Mao+Wang+2018, Cheng+etal+2020, Lan+etal+2019, Lan+etal+2020, Lan+etal+2021(2), Lan+etal+2021, Gill+etal+2020, Gill+etal+2021, Gill+Granot+2021}.

Measuring the PD from GRBs is challenging due to the deficit of gamma-ray photons, which results in a low signal-to-noise ratio. The first claimed discovery of linear polarization during the prompt phase was made in GRB 021206 \citep{Coburn+Boggs+2003}, with a PD being as high as 80$\pm$20\%, but the result was suspected to be unreliable \citep{Rutledge+Fox+2004, Wigger+etal+2004}. More reliable PD measurements are from gamma-ray burst polarimeters (GAP, \cite{Yonetoku+etal+2011}) and the Cadmium Zinc Telluride Imager (CZTI, \cite{Singh+etal+2014}). Both the instruments found that the PD of GRBs is typically higher than 50\%. Conversely, the recent measurements from POLAR revealed lower PD, with a typical value of $\lesssim10\% $ \citep{Zhang+etal+2019}. It needs to be noted that in the POLAR burst, GRB 170114A, a remarkable evolution of the PD and polarization-angle (PA) within a single pulse was observed. Specifically, the PA exhibited a significant change of approximately 90 degrees across two consecutive time bins \citep{Zhang+etal+2019}. Moreover, \citet{Burgess+etal+2019} (hereafter Burgess19) conducted a comprehensive time-resolved analysis for GRB 170114A and confirmed the evolution of PD and PA evolution. Another GRB in the POLAR catalog, GRB 170101A, was also found with a hint of PD and PA change in two time bins \citep{Kole+etal+2020}. The PA evolution was also found in the CZTI burst GRB 160821A \citep{Sharma+etal+2019} and GAP burst GRB 100826A (\citealt{Yonetoku+etal+2011, Yonetoku+etal+2012}). 

A significant consequence of the evolution of the PA is the partial cancelation of the polarization if the integrated time of the observed PD is longer than the variability timescale of the PA, resulting in the observation of diminished polarization levels. Hence, the lower PD measurements obtained by POLAR may be due to the PA evolution. If true, it suggests that the instantaneous PD of GRBs will be high, which aligns with the large-scale order MF model. We caution that the current results remain speculative since all the measurements have confidence levels below $5 \sigma$. Higher confidence level polarimetric observations from future Polarimeters, such as POLAR-2 and LEAP \citep{Kole+etal+2019, McConnel+etal+2020} are eagerly anticipated to reveal the genuine polarization properties. Interestingly, \citet{Xie+etal+2022} confirmed the substantial PD close to the synchrotron limit from the Vela pulsar wind nebula, which strongly supports the large-scale order MF configuration in this object. This provides valuable insights for studying the MF configuration in GRBs, which can also originate from compact object ejecta (e.g., magnetars or black holes). 

In theory, the PA evolution in GRB prompt emission has been expected in some models. \citet{Chang+Lin+2014} considered a magnetic-dominated jet model, in which the Lorentz factor (LF) of the jet increases with radius ($\Gamma \propto r^{1/3} $) before reaching a saturated value. If the synchrotron-Compton scattering photons are emitted before the bulk Lorentz factor saturates, the change of polarization angle is a natural result of the acceleration of the outflow.
\citet{Parsotan+etal+2020} calculated the photosphere radiation polarization and found that the PA could be changed by 90 degrees when the observer sees different portions of the jet at various times due to the structure within the jet. Moreover, the PA evolution in the afterglow emission can also be attributed to the structured jet \citep{Rossi+etal+2004}. \citet{Deng+etal+2016} found that the ICMART model \citep{Zhang+Yan+2011} will also give rise to a $90$ -degree change of the PA due to magnetic reconnection.

The PA evolution discussed above is mainly caused by jet dynamics and jet structure. In addition to these factors, the MF configuration can also cause the PA evolution. The PA evolution in the large-scale MF framework was expected in some models \citep{Cheng+etal+2020, Gill+Granot+2021, Gill+Granot+2024, Wang+Lan+2023, Wang+Lan+2023-2}. In our previous article, we considered the evolution of the instantaneous synchrotron PD and PA in the toroidal MF case \citep{Cheng+etal+2020}, which is the most possibly MF configuration in GRBs. We have found that the PA of an off-axis GRB can undergo two 90-degree rotations at most within a single pulse. This is due to the fact that the direction of the dominated MF in the visible angular region experiences two 90-degree changes with time. 
 
In this work, we carried out a more detailed investigation to explore the parameters under which the PA in a GRB pulse changes by 90 degrees, as well as the conditions that give rise to the observation of two 90-degree changes in the PA. We mainly studied the dependence of PA change on the factors of the viewing angle, the jet opening angle, and the jet Lorentz factor. 

This paper is organized as follows. In Section 2, we show the MF models and methods we used to calculate the PDs. In Section 3, we show the calculation results of the time-resolved polarization and describe three analyses in the subsections. In Section 3.1, we give the interpretation of the PD and PA evolution, especially for the explanation of the phenomenon of the PA changing by 90 degrees. In Section 3.2, we conclude on the parameter conditions of the PA changing by 90 degrees according to the results. In Section 3.3, we constrain the observed rate of GRBs with the PA changing by 90 degrees in the ordered toroidal field. In Section 4, our results are summarized and discussed.


\section{Methods and models}
\label{sec:2} 

\subsection{MF model}
We consider the large-scale ordered MF in this paper. This MF configuration generally contains two components: the poloidal and toroidal components. In the jet comoving frame,  both of the components decrease with the jet radius. The radial MF strength decreases as $B' \propto R^{-2}$, while the toroidal MF strength decreases as $B' \propto R^{-1}$ \citep{Spruit+2001}. Thus, at a large radius of the emission regions, the MF would be dominated by the toroidal component. Considering the power-law decay MF, \citet{Uhm+Zhang+2014} explained the observed GRB spectra. Here, we also generalize the decay MF strength with a pow-law form, which reads 
\begin{equation}
B'=B'_0 (\frac{R}{R_0})^{-a},
\end{equation}
where $B'_0$ is the initial MF strength, $R_0$ is the radius where the MF begins to decay, $a$ is the MF decaying index, and $R=R_s+\beta c\Gamma t'$ is the jet radius. $R_s$ is the radius where the GRB emission starts, $\Gamma$ is the bulk Lorentz factor of the jet, and $\beta$ is the dimensionless velocity of the jet. For simplicity, we only considered $a=-1$ in this work, i.e., the MF is toroidal component dominated.

\subsection{The flux-density spectra}
The observed flux density of synchrotron radiation from an electron population with a distribution of $dN'_e/d\gamma'_e$ can be calculated as (e.g., \citealt{Geng+etal+2018})
\begin{equation} \label{continu_equ}
F_\nu \approx \frac{\Gamma(1+z)}{4\pi D_L^2}\int^{\gamma'_{\rm{max}}}_{\gamma'_m} P'(\gamma'_e,\nu')\frac{dN'_e}{d\gamma'_e}d\gamma'_e.
\end{equation}
where $\gamma'_m$ and $\gamma'_{\rm{max}}$ are the minimum and maximum electron LFs, respectively. $z$ is the redshift, and $D_L$ is the luminosity distance. $P'(\gamma'_e,\nu')$ is the spectral power of the synchrotron radiation from a single electron in the comoving frame \citep{Rybicki+Lightman+1979}:
\begin{equation}
P'(\gamma'_e,\nu')=\frac{\sqrt{3}q_{e}^3 B' \sin\alpha'}{m_e c^2}[x\int_{x}^{\infty} K_{5/3}(\xi)d\xi],
\end{equation}
where $x=\nu'/\nu'_{c}$, $\nu'_c=\frac{3q_{e}B'\sin\alpha'}{4\pi m_e c}\gamma_{e}^{'2}$, $K_{5/3}(\xi)$ is the Bessel function, and
$\alpha'$ is the angle between the MF and the electron velocity. $q_e$ and $\gamma'_{e}$ are the electron charge and electron Lorentz factor, respectively. The electron distribution $dN'_e/d\gamma'_e$ can be obtained by solving the continuity equation of the electrons in energy space (e.g., \citealt{Longair+2011,Zhao+etal+2014}):
\begin{equation} \label{continu_equ}
\frac{\partial}{\partial t'}\left(\frac{dN'_e}{d\gamma'_e}\right) + \frac{\partial}{\partial \gamma'_e}
\left[\dot{\gamma}'_{e,tot}\left(\frac{dN'_e}{d\gamma'_e}\right)\right] = {Q'}(\gamma'_e)
,\end{equation}
where $\dot{\gamma}'_{e,\rm{tot}}$ is the total cooling rate of the electrons, and $Q'(\gamma'_e)=Q'_{0}\gamma_{e}^{\prime-p} $ (for $\gamma'_{m}<\gamma'_{e}<\gamma'_{\rm{max}} $) is
the electron injection term, which is assumed to be a power-law form. $p$ is the electron index. The cooling mechanisms of high-energy electrons in GRB emission mainly contain synchrotron (SYN), adiabatic (ADI), and synchrotron self-Compton (SSC) coolings. For simplicity, we only consider the SYN and ADI coolings. Thus, the total cooling rate of the electrons can be obtained as $\dot{\gamma}'_{e,\rm{tot}}=\dot{\gamma}'_{e,\rm{syn}}+\dot{\gamma}'_{e,\rm{adi}}
$. The SYN cooling rate is given by \citep{Rybicki+Lightman+1979}
\begin{equation}
\dot{\gamma}'_{e,\rm{syn}}=-\frac{\sigma_{T} B^{'2} \gamma_{e}^{'2}}{6 \pi m_{e} c}.
\end{equation}
The ADI cooling rate is (e.g., \citealt{Tavecchio+etal+2003, Uhm+etal+2012})
\begin{equation}
\dot{\gamma}'_{e,\rm{adi}}=-\frac{2}{3} \frac{\gamma'_{e}}{R} \frac{dR}{dt'}=-\frac{2}{3} \frac{\beta c \Gamma \gamma'_{e}}{R}.
\end{equation}
\subsection{The calculation of synchrotron PD}
\label{subsec:2.3}
For a given observed wavebands [$ \nu_{1} , \nu_{2} $], the linear PD of GRB prompt emission can be calculated as \citep{Cheng+etal+2020}
\begin{equation}\label{pi}
\begin{aligned}
\Pi &=\int_{\nu_1}^{\nu_2}d\nu \int_{0}^{(1+q)^{2}y_{j}} f(y) dy \int_{-\Delta \phi(y)}^{\Delta \phi(y)} d\phi  \int_{\gamma'_{\rm{m}}}^{\gamma'_{\rm{max}}} G(x)\frac{dN_{e}}{d\gamma'_e}(t') \\
&\times B(t') \sin \alpha' \cos (2\chi ) d\gamma'_{e}  [\int_{\nu_1}^{\nu_2}d\nu \int_{0}^{(1+q)^{2}y_{j}} f(y) dy\\
&\times \int_{-\Delta \phi(y)}^{\Delta \phi(y)} d\phi  \int_{\gamma'_{\rm{m}}}^{\gamma'_{\rm{max}}} F(x)\frac{dN_{e}}{d\gamma'_e} (t') B(t')  \sin \alpha'  d\gamma'_{e} ]^{-1}
\end{aligned}
,\end{equation}
where $t'= t_{\rm{obs}}\delta/(1+z)$ is the jet comoving time and $t_{\rm{obs}}$ is the observed time. We define several variables here, $y \equiv (\Gamma \theta)^{2}$,  $y_{j} \equiv (\Gamma
\theta_{j})^{2}$, and $q \equiv \theta_{v} / \theta_{j}$, where  $\theta_{v}$ is the viewing angle.
$f$ is a function of $y$. For the instantaneous polarization, $f(y)=(1+y)^{-3}$ (Nakar \& Piran 2003). $F(x)$ and $G(x)$ can be written as \citep{Rybicki+Lightman+1979}
\begin{equation}
\begin{cases}
F(x)=x\int_{x}^{\infty}K_{5/3}(\xi)d\xi \\
G(x)=xK_{2/3}(x)
\end{cases}
,\end{equation}
where $K_{5/3}(\xi)$ and $K_{2/3}(x)$  are Bessel functions. Other variables are written as \citep{Granot+2003, Granot+konigl+2003, Granot+Taylor+2005, Toma+etal+2009}
\begin{equation}
\sin \alpha' = \left[ \left(\frac{1-y}{1+y}\right)^{2} + \frac{4y}{(1+y)^{2}} \frac{(s-\cos \phi)^{2}}{(1+s^{2}-2s \cos \phi)} \right]^{1/2}
,\end{equation}
\begin{equation}
\chi = \phi + \rm{arctan} \left( \frac{(1-y)}{(1+y)} \frac{\sin \phi}{(a- \cos \phi)}\right)
,\end{equation}
\begin{equation}
\Delta \phi(y) =
\begin{cases}
0,    \qquad \qquad \qquad \qquad \rm{for} \; q>1  \; \rm{and}  \;  y<(1-q)^2 y_{j}     \\
\pi,  \qquad \qquad \qquad \qquad \rm{for} \; q<1  \; \rm{and}  \;  y<(1-q)^2 y_{j}   \\
\cos^{-1} \left[\frac{(q^{2}-1)y_{j}+y}{2q \sqrt{y_{j}y}}\right]   \qquad \qquad \qquad \qquad \quad \rm{otherwise}.
\end{cases}
,\end{equation}
where $s=\theta/\theta_{v}$. We assume the emission generates from a thin shell with a radius range of $R_s$ to $R_{\rm{off}}$, where $R_s$ is the radius where emission starts and $R_{\rm{off}}$ is the radius where emission turns off. The starting times of the emission are $t_{\rm{obs0}}=R_{s}(1+z)[1-\beta]/(\beta c)$ and $t_{\rm{obs0}}=R_{s}(1+z)[1-\beta \cos(\theta_{v}-\theta_{j})]/(\beta c)$ for the on-beaming and off-beaming cases, respectively. The turn-off times of the emission are $t_{\rm{off}}=R_{\rm{off}}(1+z)[1-\beta]/(\beta c)$ and $t_{\rm{off}}=R_{\rm{off}}(1+z)[1-\beta \cos(\theta_{v}-\theta_{j})]/(\beta c)$ for the on-beaming and off-beaming cases, respectively. We note that we set $R_{\rm{off}}=3R_s$ in this paper; thus, we have $t_{\rm{off}}=3 t_{\rm{obs0}}$. We adopt a set of typical parameters in our calculations: $R_0=R_s =1 \times 10^{14} $cm, $p=2.8$, $B'_0=2000$G, $\gamma'_m =1\times 10^{4}$, $\Gamma=300$, $\theta_j =0.1$ radians, $D_L=1\times 10^{28}$cm, and $z=1$, where $B'_0$, $\gamma'_m$, and $\Gamma$ are constrained with the observed spectral peak energy $E_{\rm{peak}}=1/(1+z)\cdot 3hq_eB'_0\Gamma\gamma_m^{\prime 2}/4\pi m_e c\sim 500 \rm{keV}$.

\section{Instantaneous polarization properties}
\label{sec:3}

In this work, we concentrated on investigating the evolution of PD and PA during gamma-ray burst (GRB) prompt emission, with a specific focus on elucidating the underlying physical mechanisms and the parameter conditions under which PA undergoes a 90-degree change.
We calculate the instantaneous PDs in three wavebands, including the optical ($4.3\times 10^{14} - 7.5\times 10^{14} $Hz), X-ray (2-30 keV), and $\gamma$-ray bands (30-800 keV). The optical band is within the detected wavebands of the optical polarimeter RINGO3 \citep{Arnold+etal+2012}. The X-ray and $\gamma$-ray bands are the energy ranges of the Low energy Polarimetry Detector \citep{Fan+etal+2023} (LPD, 2-30 keV) and the High energy Polarimetry Detector (HPD, 30-800 keV) in POLAR-2, respectively. 

\subsection{The evolution of instantaneous polarization and its interpretation}
\label{subsec:3.1}

\begin{figure*}
   \centering
   \includegraphics[width=\textwidth]{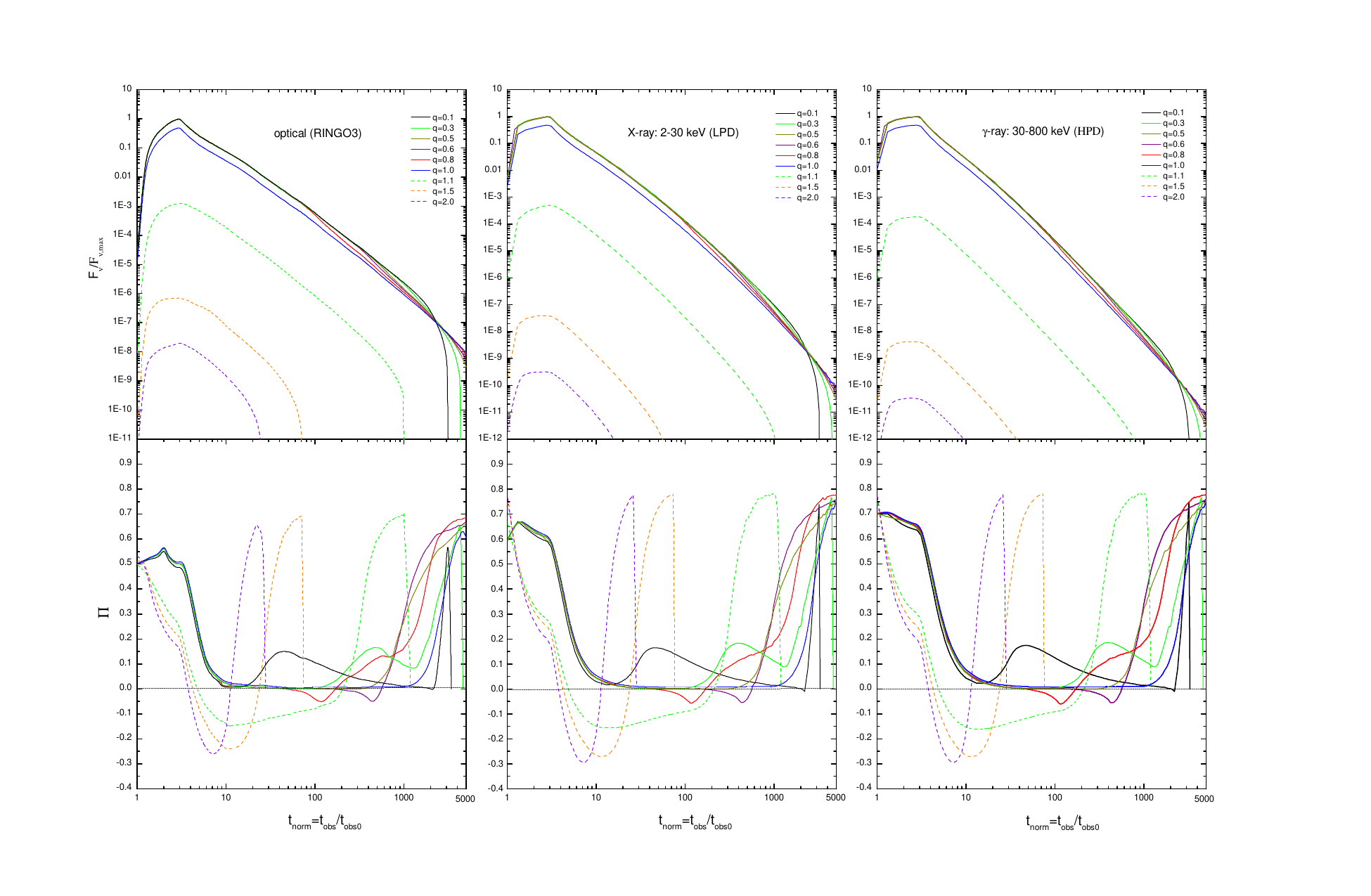}
      \caption{Instantaneous PDs with different viewing angles (bottom panels) and corresponding normalized light curves (upper panels, normalized to the maximum flux). The left, middle, and right panels are calculated in the optical ($4.3\times 10^{14} - 7.5\times 10^{14} $Hz), X-ray (2-30 keV), and $\gamma$-ray (30-800 keV) bands, respectively. We note that $t_{\rm{obs0}}$ is the starting time of the GRB pulse and $t_{\rm{norm}}$ is the time normalized to the starting time. }
         \label{insta_pola_3_bands}
   \end{figure*}

Figure \ref{insta_pola_3_bands} shows the instantaneous PDs and the corresponding normalized flux with different viewing angles in the three wavebands. We note that all of the normalized fluxes in this paper are normalized to the same on-axis peak flux with the set of typical parameters. The instantaneous PDs for the optical, X-ray, and gamma-ray bands have minor differences. All of the polarization curves for $q>0.5$ ($q=0.6,0.8,1.1,1.5$, and $2.0$) decline rapidly from large positive PDs at early times to negative ones, and then they experience another turnover, returning to positive values again at late times. This means that the PA experiences two changes by 90 degrees. On the other hand, the PAs for the cases of $q \leq 0.5$ remain unchanged, while the PDs decrease from initial positive values to nearly zero, then rise to a peak, decrease again to zero, and eventually return to their initial values. The results of the polarization curves for $q>0.5$ are consistent with those of \citet{Gill+Granot+2021} (hereafter GG21); but, comparatively, we add the calculations of $q \leq 0.5$, and the results obtained were significantly different from those with $q>0.5$, which is mainly reflected by whether the PA undergoes a 90-degree reversal.

\begin{figure*}
   \centering
   \includegraphics[width=\textwidth]{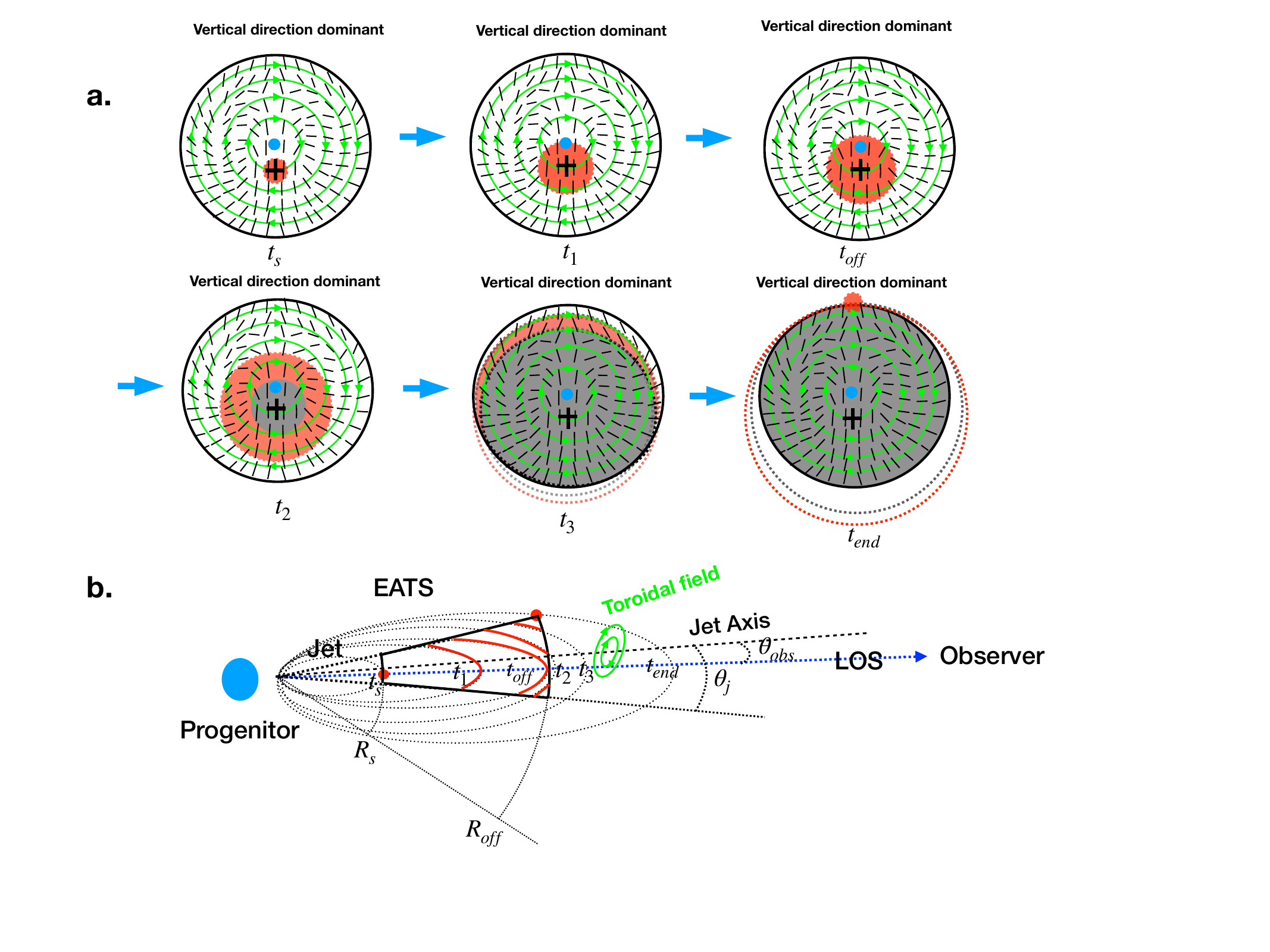}
      \caption{
Schematic diagram of PA evolution for case of $q<0.5$.
The upper panels (a.) show the change of the observed jet area with time on the plane of the sky. The green circle with an arrow represents the toroidal MF. The black plus symbol represents the LOS. The red area shows the observed radiation region. The gray area shows where radiation has disappeared. The short dotted lines represent the directions of the dominated polarization vectors. The bottom panel (b.) shows the change of observed jet area with time on the lateral section. The red solid lines and points represent the surfaces formed by the intersection of EATS and jet at different times.   }
         \label{schema_on_beam}
   \end{figure*}

\begin{figure*}
   \centering
   \includegraphics[width=\textwidth]{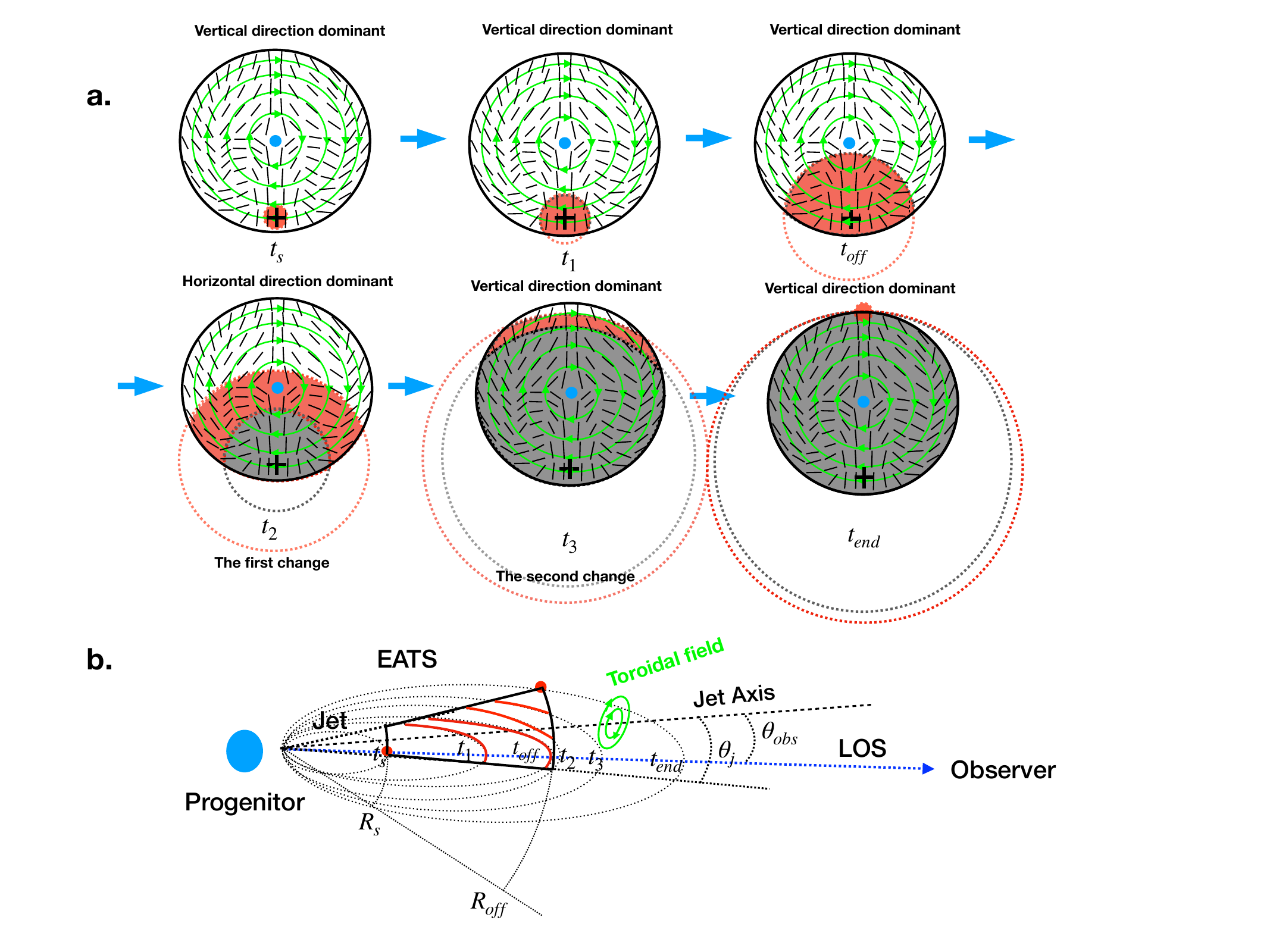}
      \caption{  
Similar to Fig. \ref{schema_on_beam}, but with schematic diagram of PA evolution for the case of $q>0.5$.}
         \label{schema_off_beam}
   \end{figure*}

These polarization properties can be explained by referring to Figs. \ref{schema_on_beam} and \ref{schema_off_beam}.
Fig. \ref{schema_on_beam} is the schematic diagram of the polarization evolution for the case of $q <0.5$. The upper panels (a.) depict the variation of the observed jet area over time in the plane of the sky. The bottom panel (b.) shows the change of observed jet area over time in the lateral section. The red solid lines and points represent the surface formed by the intersection of the equal-arrival-time surface (EATS) and jet for different times. 

\begin{figure*}
   \centering
   \includegraphics[width=\textwidth]{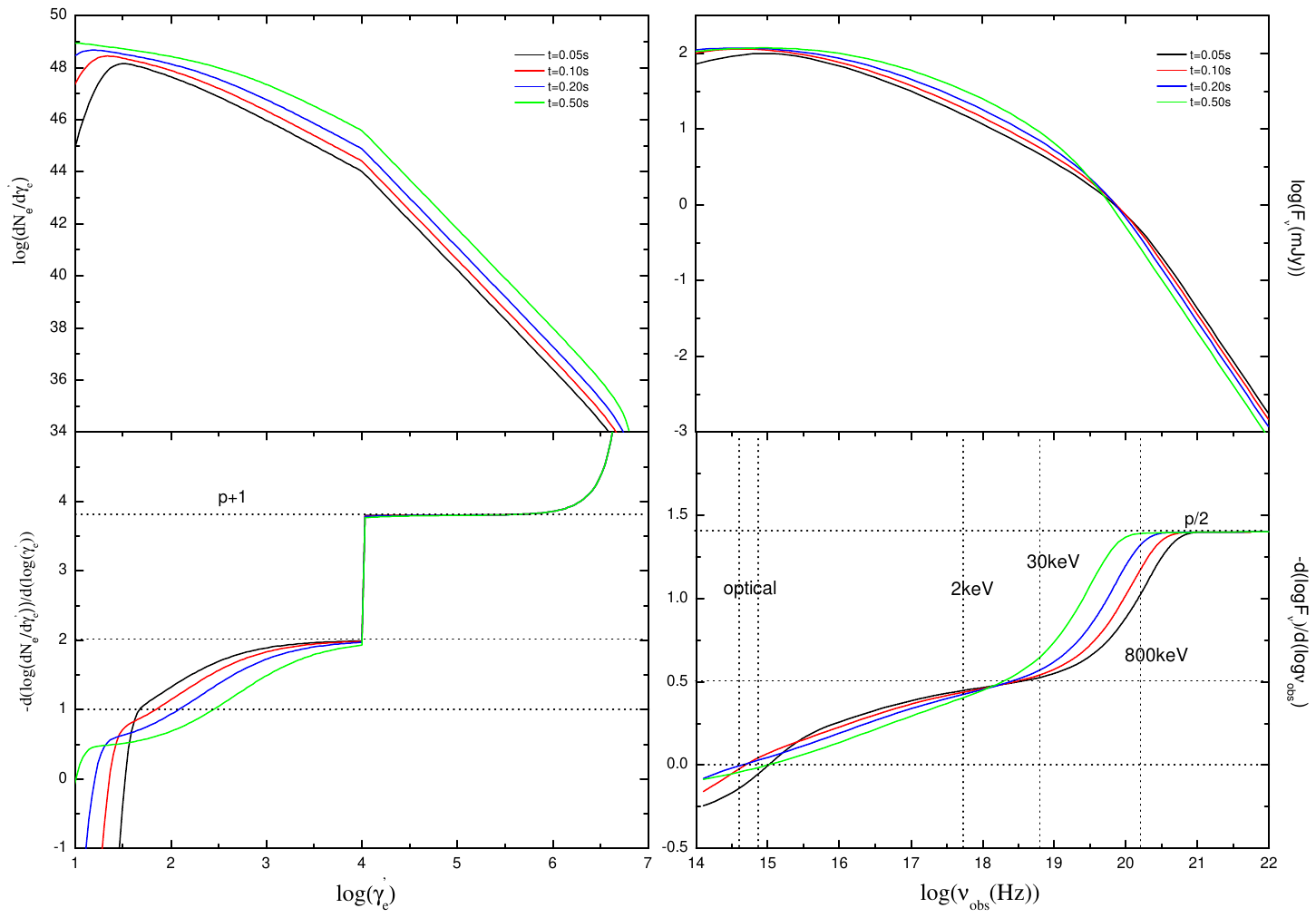}
      \caption{Evolution of electron distribution (the upper left panel) and corresponding flux-density spectra (the upper right panel) for the on-beaming case in our polarization calculation. The bottom panels show the negative spectral index of the electron distribution (the bottom left panel) and flux-density spectra (the bottom right panel), respectively. The cases in the optical, X-ray, and $\gamma$-ray bands are marked in the bottom right panel, respectively.
                           }
         \label{ele_spec_slope}
   \end{figure*}

As is shown in Fig. \ref{schema_on_beam}, when the observer receives photons at early times (around $t_s$), the observed area is just a point-like zone. The polarization thus is approximately the intrinsic polarization, i.e., the PD is as high as $(\alpha+1)/(\alpha+5/3)$ \citep{Rybicki+Lightman+1979}, where $\alpha$ is the spectral slope. In the optical, X-ray, and $\gamma$-ray bands and for the on-beaming case ($q \lesssim 1$), the initial PDs are $\sim 0.5$, $\sim 0.6$, and $\sim 0.7$, respectively. These PDs can be easily estimated from the corresponding flux density spectra in Fig. \ref{ele_spec_slope}, where distinct spectral indices ($\alpha$) for the three bands can be discerned from the bottom right panel of the figure.

The observable area begins to extend from $t_s$ to $t_1$ and then to $t_{\rm{off}}$ (see Fig. \ref{schema_on_beam}), where $t_{\rm{off}}$ is the time when photons from the turn-off radius $R_{\rm{off}}$ begin to reach the observer 
and $t_1$ is a time between $t_s$ and $t_{\rm{off}}$. The polarization vectors are dominated by the vertical direction. After $t_{\rm{peak}} (\sim t_{\rm{off}})$, the region where the radiation ceases spreads (gray region in Fig. \ref{schema_on_beam}), while the observable region (red region in Fig. \ref{schema_on_beam}) gradually extends toward high latitudes. However, the polarization vectors are still dominated by the vertical direction between $t_2$ and $t_{\rm{end}}$. Hence, the PA remains unchanged. 

The results shown in Fig. \ref{schema_off_beam} correspond to the case of $q>0.5$. Similarly to Fig. \ref{schema_on_beam}, due to the beaming effect, the radiation would be dominated by the areas closer to the LOS. Thus, the polarization vectors are always dominated by the vertical direction for the time of $t_s - t_{\rm{off}}$. As a result, the PA remains unchanged until $t_{\rm{off}}$. The difference compared to the case of $q<0.5$ is that after $t_{\rm{off}}$, the direction of the polarization vectors from high latitudes changes from the vertical dominant to the horizontal dominant at $t_2$. The corresponding PA changes by 90 degrees for the first time, followed by another change up to $t_3$. At $t_2$, the viewing field is composed of two parts: the observed radiation region (red area) and the radiation disappearance region (gray area). The polarization vector of the radiation disappearance region is vertically dominant, while the remaining radiation region is dominated by the horizontal direction. Thus, the observed radiation is dominated by the horizontal direction. This causes a 90-degree reversal of the polarization direction compared to before $t_{\rm{off}}$. At $t_3$, it experiences a similar reversal. At the end of radiation ($t_{\rm{end}}$), the observer can only capture the radiation generated by a point-like zone at the highest latitude. Consequently, the PD is approximately restored to its intrinsic value. The PA thus experiences two 90-degree variations. Hence, we can conclude two reasons for the PA flip in GRB prompt emission as follows. One is the low level of symmetry of the observed area about the direction of the LOS compared to the case of $q<0.5$. The other one is the radiation disappearance in local regions in the viewing field, which could change the dominant direction of the polarization vectors. Compared with the explanation of the 90-degree PA flip in GG21, we divide it into two situations -$q>0.5$ and $q < 0.5-$ based on whether or not the PA flip will occur. 

We should note that $q \simeq 1.0$ is a special case, and there are no 90-degree PA changes for this case. The PD gradually approaches zero after the initial decline, and it remains at zero until the late period before it rises again (see Fig. \ref{insta_pola_3_bands}). It means that the vertical and horizontal polarization vectors just cancel each other out during the long period of PD $\sim 0$. This long period is between $t_{\rm{off}}$ and $t_3$ in Fig. \ref{schema_off_beam}, and the panel of $t_2$ in Fig. \ref{schema_off_beam} is no longer applicable to explain this case. When the LOS is along the edge of the jet, the vertical and horizontal polarization vectors just cancel each other out in the red regions. After $t_3$, the polarization vectors would be dominated by the vertical direction, and the PD will rise again.

In order to compare the detailed difference of the PD evolution across different wavebands and their relation with light curves, we illustrate the polarization curves and light curves for the three wavebands in Fig. \ref{flux_pola_t}. As depicted in the bottom panel of the figure, after the pulse's peak time, there are slight differences in the PD evolution among the three wavebands. This suggests that the PD evolution in the decay phase of the pulse is predominantly determined by the geometrical effect. Furthermore, it can be seen that the decay of the light curve in the optical band is much more gradual compared with those in the X-ray and $\gamma$-ray bands. For $q=0.8$ and take the set of typical parameters, the fluxes corresponding to the times when the first and second PA change by 90 degrees are around $10^{-8}$ erg $\rm{s^{-1} cm^{-2}}$ and $10^{-10}$ erg $\rm{s^{-1} cm^{-2}}$ in the X-ray or $\gamma$-ray bands. Thus, the fluxes change by roughly two orders of magnitude between the first PA change, and the second PA changes in the two bands, while the decrease is only one order of magnitude in the optical band (from $\sim 10^{-11}$ to $\sim 10^{-12}$ erg $\rm{s^{-1} cm^{-2}}$). This means the second PA change is more readily observable in the optical band. 

\subsection{The parameter conditions of the 90-degree PA flip}
\label{subsec:3.2}
\begin{figure*}
   \centering
   \includegraphics[width=\textwidth]{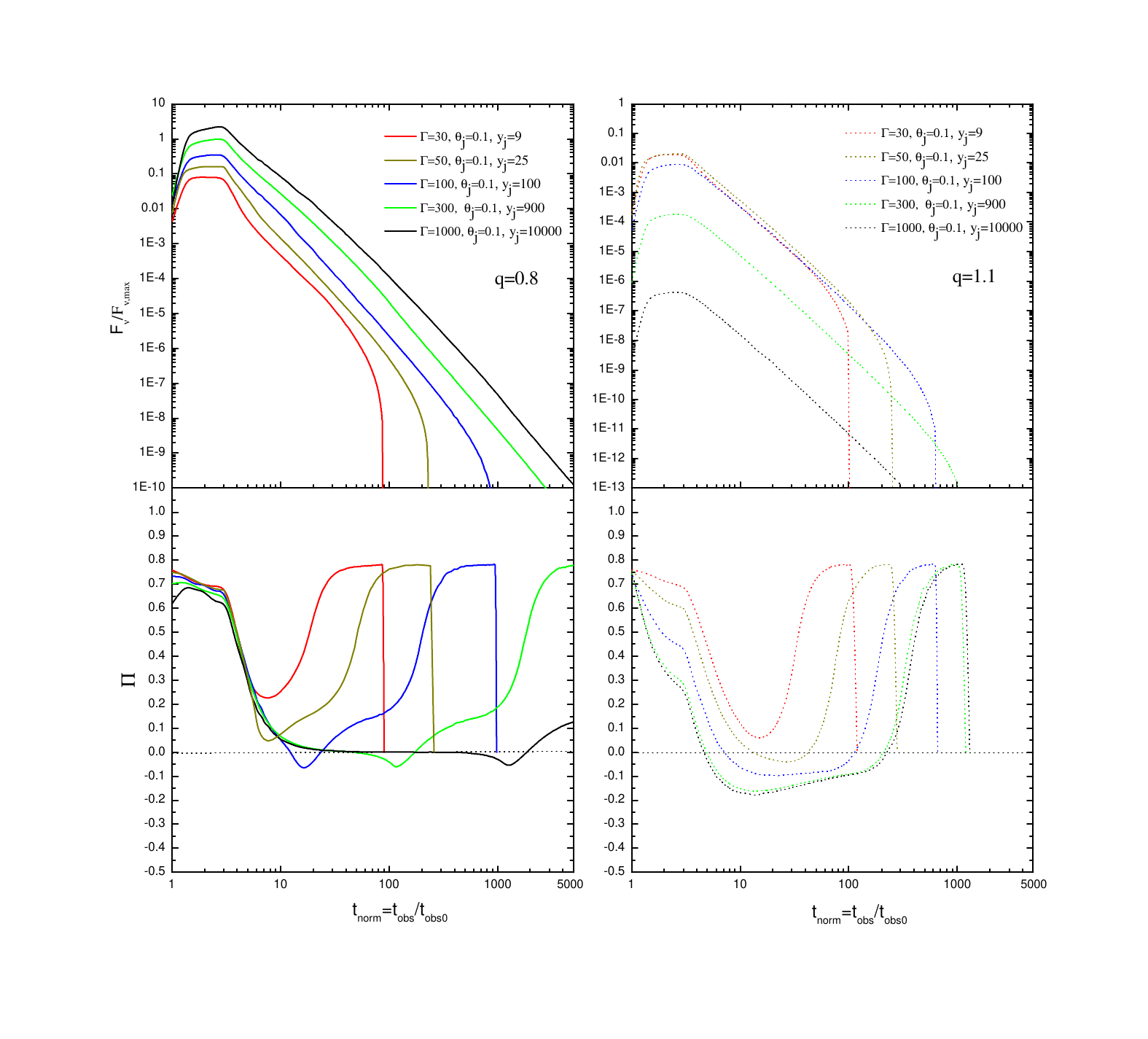}
      \caption{ Instantaneous polarization curves with different values of $y_j $. The jet opening angle is fixed at $\theta_j =0.1,$ and the jet LF is variable.
                           }
         \label{polar_flux_with_LZ}
   \end{figure*}

\begin{figure*}
   \centering
   \includegraphics[width=\textwidth]{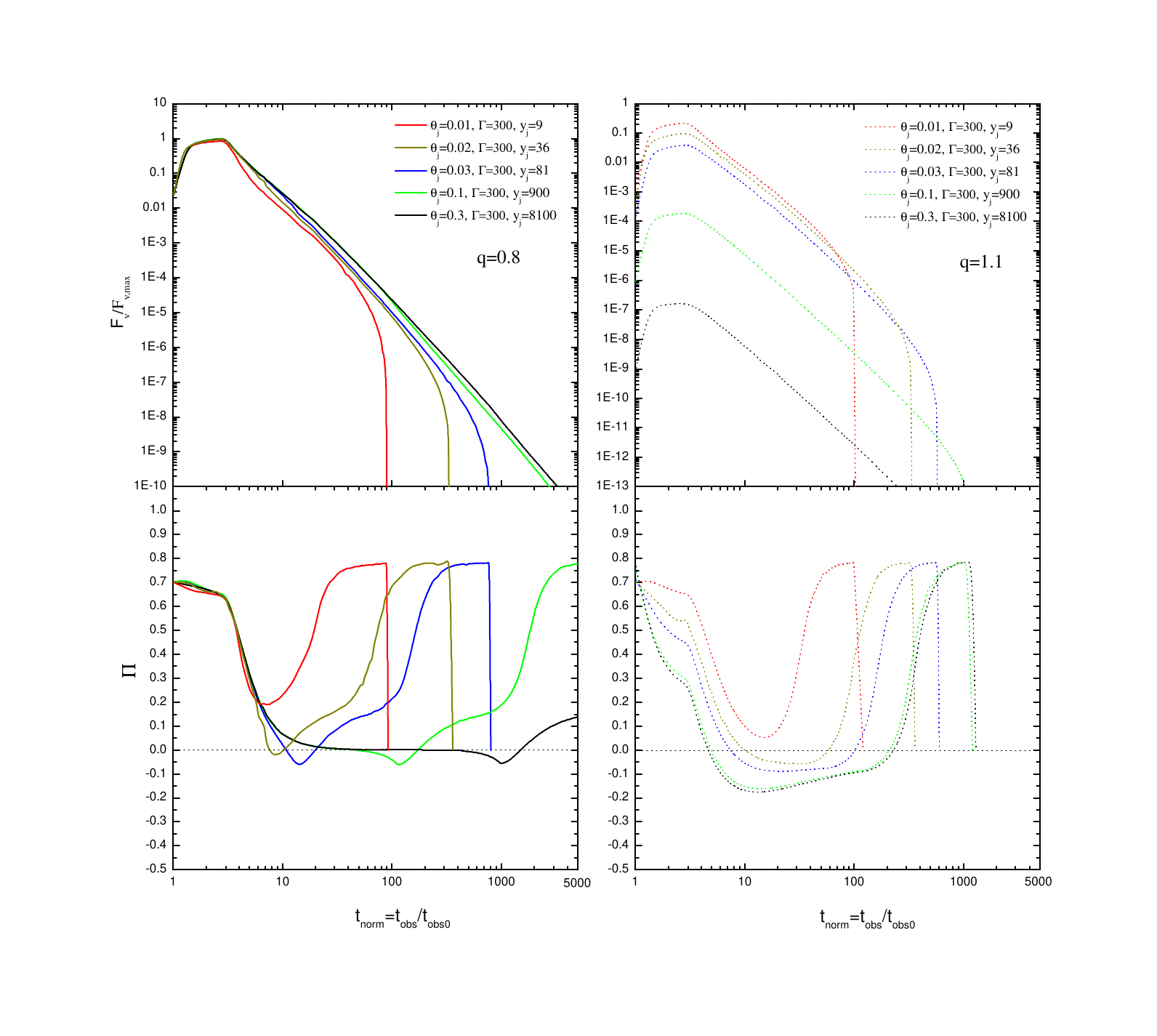}
      \caption{Similar to Fig. 4, but invariant is the jet LF. The LF is fixed at $\Gamma =300,$ and the jet opening angle is variable.     
                           }
         \label{polar_flux_with_thj}
   \end{figure*}

As discussed in Section \ref{subsec:3.1}, the PA could change by 90 degrees in GRB prompt emission in a pulse. In order to investigate the conditions under which this PA change occurs, we calculated the instantaneous PDs from different viewing angles, jet LFs, and jet half-opening angles. We find that the 90-degree PA change should be satisfied with $q \gtrsim 0.5$ from Fig. \ref{insta_pola_3_bands}. The instantaneous PDs under different jet LFs and jet half-opening angles are shown in Figs. \ref{polar_flux_with_LZ} and \ref{polar_flux_with_thj}, respectively.
Both figures show the polarization curves with different values of $y_j=(\Gamma \theta_j)^2$. In Fig. \ref{polar_flux_with_LZ}, we fix the jet half-opening angle to $\theta_j =0.1$ and then change the jet LFs. Next, we fix the jet LFs to $\Gamma=300$ and then change $\theta_j$ in Fig. \ref{polar_flux_with_thj}. We calculated the polarization curves in cases of $q=0.8$ and $1.1$. As shown in Figs. \ref{polar_flux_with_LZ} and \ref{polar_flux_with_thj},  
the PA does change by 90 degrees for $y_j \gtrsim 36$. Hence, we can roughly conclude that $y_j \gtrsim 36 $ is a condition for the PA to change by 90 degrees. To refine the parameter conditions for the PA flip, we explore the two-dimensional parameter space of $y_j$ and $q$ in relation to the normalized flux of the PA flip.

\begin{figure*}
   \centering
   \includegraphics[width=\textwidth]{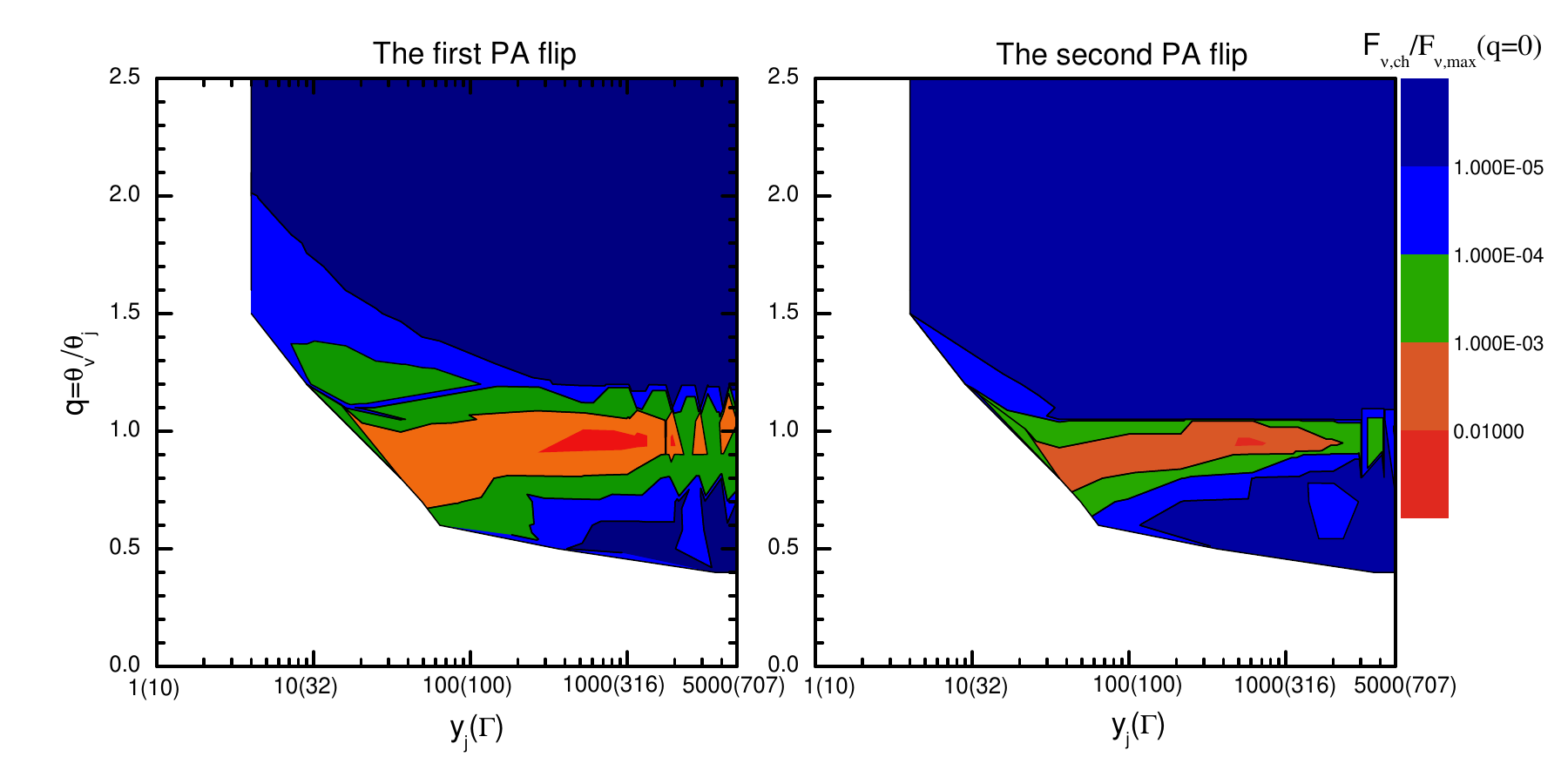}
      \caption{Contour plots of normalized flux at the time of the first (left panel) and second (right panel) PA flip in the two-dimensional parameter space of $q$ versus $y_j$. We note that there is no PA flip when $q\simeq1$, which is not marked in the figure. In this group of calculations, we fixed $\theta_j=0.1$ radians and changed $\Gamma$. The numbers in parentheses on the horizontal axis represent the values of the corresponding Lorentz factors.
      }
         \label{contour-nflux}
   \end{figure*}

We map the normalized flux at the time of the first and second PA flip in the two-dimensional parameter space of $q$ versus $y_j$. The results are shown in Fig. \ref{contour-nflux}. In this figure, we fixed $\theta_j=0.1$ radians and then changed $\Gamma$ to change the value of $y_j$. According to this figure, we can obtain the parameter conditions of PA flip as $q\gtrsim 0.5$ (except for $q\simeq 1$) and $y_j \gtrsim 4$.
Moreover, as the LOS approaches the jet edge, the normalized flux at the PA flip time increases, reaching its maximum of $10^{-4}$ for $0.9\lesssim q <1$. 
Hence, it is easier to observe the PA flip when observing around the edge of the jet.

\begin{figure*}
   \centering
   \includegraphics[width=\textwidth]{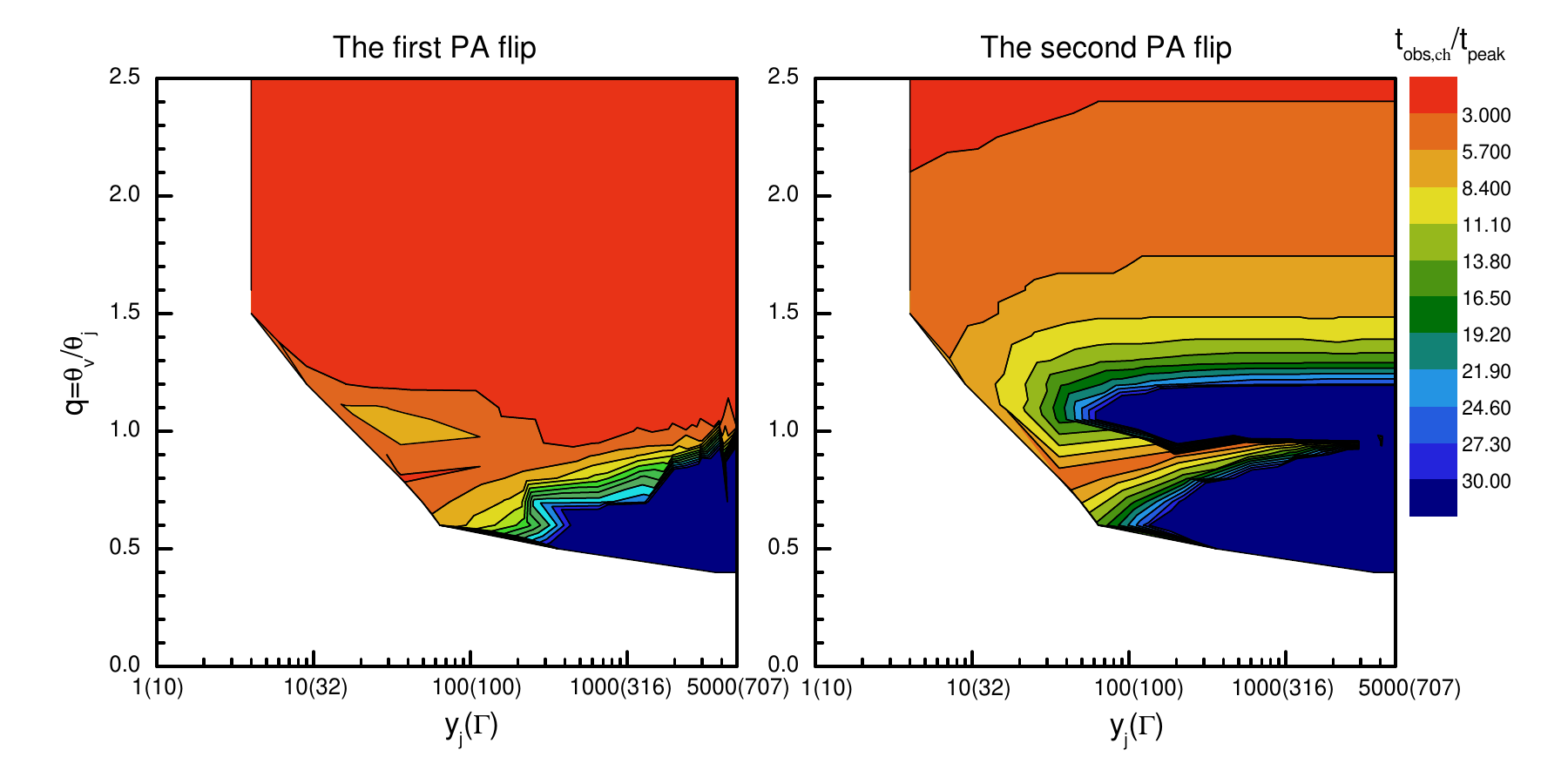}
      \caption{Contour plots of normalized time (normalized to the peak time) of the first (left panel) and second (the right panel) PA flips in two-dimensional parameter space of $q$ versus $y_j$.    
                           }
         \label{contour-nflux-t}
   \end{figure*}

In order to determine the time nodes of PA flips, we also map the time ($t_{\rm{obs,ch}}/t_{\rm{peak}}$, normalized to the peak time) when PA flips in the two-dimensional parameter space of $q$ versus $y_j$. The contour plots are shown in Fig. \ref{contour-nflux-t}. For the off-beaming ($q\gtrsim 1$) case, the larger the value of $q$, the earlier the PA flip occurs. The time of the PA flip could be earlier than $3t_{\rm{peak}}$. For the on-beaming ($q\lesssim 1$) case, the earliest time for PA flip occurs when the value of $q$ is in the range of $\sim (0.8-0.9)$.

\subsection{Estimation of the observed rate of GRBs with the 90-degree PA flip}
\label{subsec:3.3}
\begin{figure*}
   \centering
   \includegraphics[width=\textwidth]{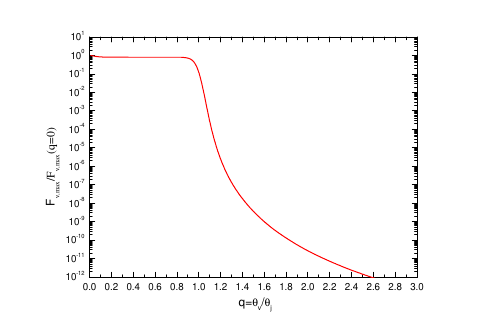}
      \caption{Ratio of maximum flux to maximum flux of $q=0$ with different viewing angles (different $q$ values)
                           }
         \label{F-q}
   \end{figure*}

We give the parameter conditions of a 90-degree PA flip as $q\gtrsim 0.5$ (except for $q\simeq 1$) and $y_j \gtrsim 4$ in Section \ref{subsec:3.2}. Following these conditions, we can limit the observed rate of GRBs with PA change to 90 degrees. Given the rapid drop of the flux when $q > 1$ (see Fig. \ref{F-q}), the case for the GRBs with $q > 1.1$ is neglected. Moreover, considering the condition of $q \gtrsim 0.5$ and assuming the orientation between the jet axis and the LOS is random, we can give a theoretical upper limit of the observed rate of GRBs with a 90-degree PA flip (including single and double flips) in the $\gamma$-ray band as 
\begin{equation}
\label{12}
R_{ch} \lesssim R_{lim} \simeq \frac{\Omega_{lim}}{\Omega_{tot}} \simeq \frac{\cos{(0.5 \theta_j)}- \cos{(1.1\theta_j)}}{1-\cos{(1.1 \theta_j)}}
,\end{equation}
where $\Omega_{lim}$ and $\Omega_{tot}$ are the solid angle of the viewing angles within $[0.5\theta_j, 1.1\theta_j]$ and $[0, 1.1\theta_j]$, respectively, and can be calculated as
\begin{equation}
\begin{cases}
\Omega_{lim}\simeq \int_{0.5\theta_j}^{1.1\theta_j} \sin{\theta} d\theta \int_{0}^{2\pi} d\phi =2\pi [\cos{(0.5 \theta_j)}- \cos{(1.1\theta_j)}] \\
\Omega_{tot}\simeq\int_{0}^{1.1\theta_j} \sin{\theta} d\theta \int_{0}^{2\pi} d\phi =2\pi [1-\cos{1.1\theta_j}]
\end{cases}
.\end{equation}
Based on Equation \ref{12}, the maximum observed rate of GRBs with a PA change of 90 degrees is the function of $\theta_j$. Taking the typical jet opening angle $\theta_j \sim 0.1$, we can constrain the observed rate as $R_{ch} \lesssim R_{lim} \simeq \frac{\cos{(0.5 \times 0.1)}- \cos{(1.1 \times 0.1)}}{1-\cos{(1.1\times 0.1)}}\sim 80 \% $. This is the upper limit of the observed rate; to estimate the observed rate we also need to make the following calculations.
We calculated the parameter space of the normalized flux at the time of a 90-degree PA flip in Section \ref{subsec:3.2}, but measuring such a PA flip is not directly determined by the normalized flux at the time of PA flip, the polarization (PD and PA) must be measured both before and after this PA flip. Considering the much lower flux and PD level in the latter case, it would be more difficult to measure the polarization after the PA flip. Thus the measure of the PA flip is mainly determined by the measure after the PA flip. Moreover, the key to detecting the polarization lies in measuring the polarized flux. Therefore, the key to measuring such a PA flip is to measure the polarized flux after this PA flip. Thus, studying the distribution of polarized flux after the PA flip in the parameter space would be useful in order to indicate the observed rate of GRBs with such a PA flip. 

\begin{figure*}
   \centering
   \includegraphics[width=\textwidth]{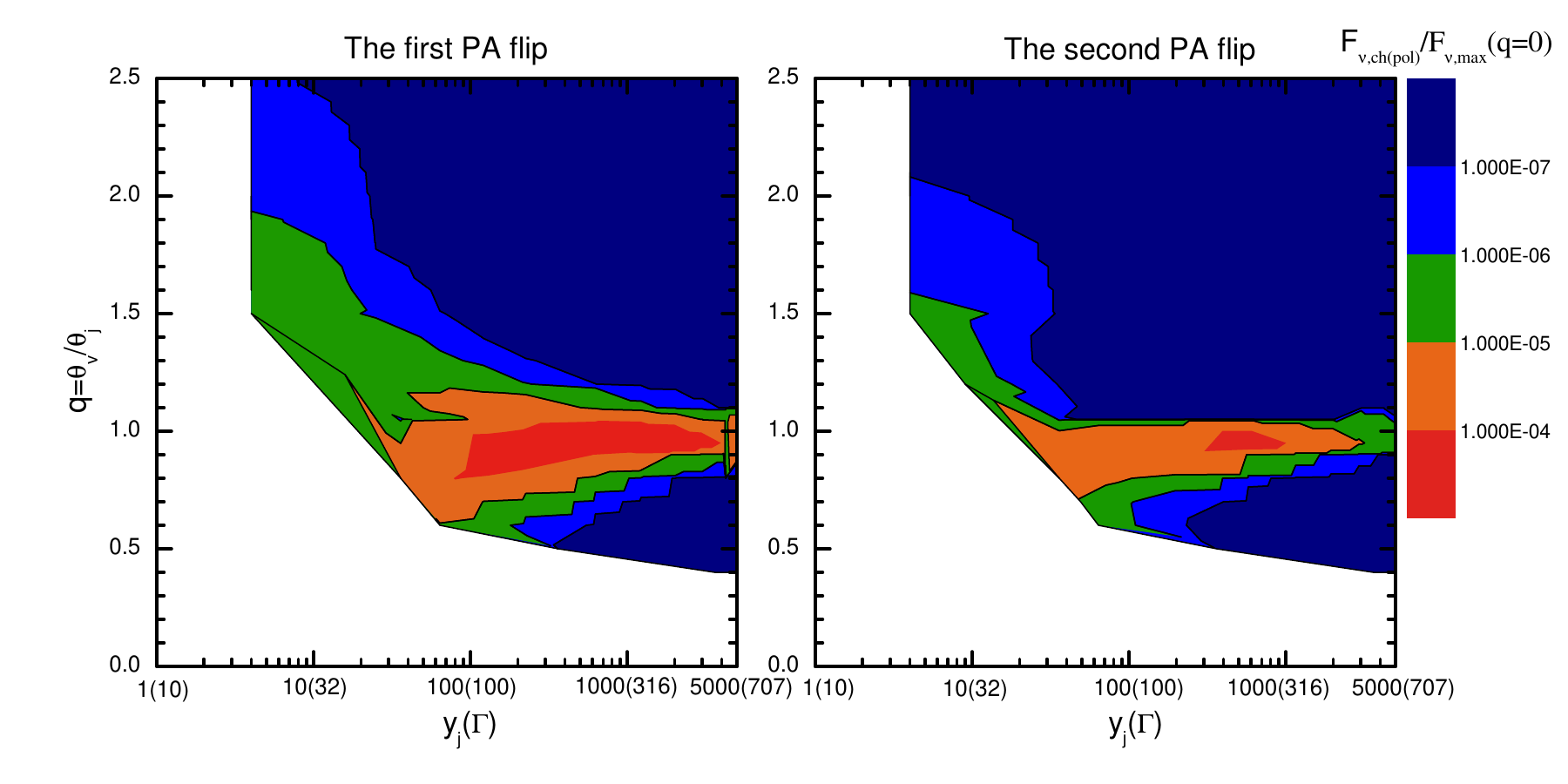}
      \caption{Contour plots of maximal post-flip polarized flux of first (left panel) and second (right panel) PA flip in the two-dimensional parameter space of $q$ versus $y_j$. In this group of calculations, we set $\theta_j=0.1$ radians and change $\Gamma$.  
                           }
         \label{contour-cgama}
   \end{figure*}

\begin{figure*}
   \centering
   \includegraphics[width=\textwidth]{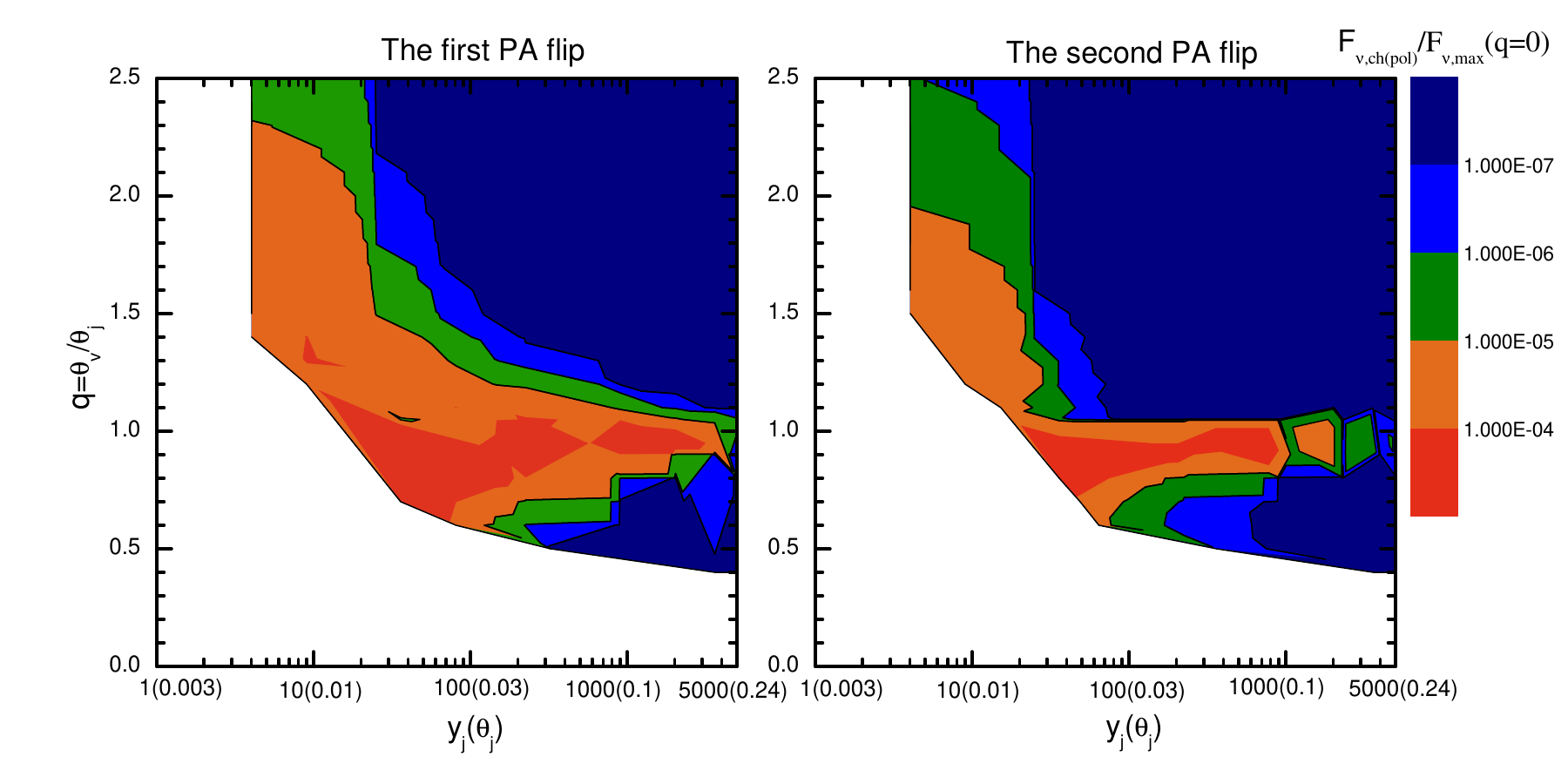}
      \caption{Similar to Fig. 11, but we set $\Gamma=300$ and change $\theta_j$ in this group of calculations.     
                           }
         \label{contour-cthj}
   \end{figure*}

In order to indicate the observed rate of GRBs with PA flip, we map the maximal post-flip polarized flux (normalized to on-axis peak flux) in a two-dimensional parameter space of $q$ versus $y_j$. The contour plots are shown in Fig. \ref{contour-cgama} and Fig. \ref{contour-cthj}. In Fig. \ref{contour-cgama}, $\theta_j$ is fixed at $0.1$ radians and then changed,  $\Gamma$. In Fig. \ref{contour-cthj}, $\Gamma$ is fixed at $300$ and then changed, $\theta_j$. As shown in these two figures, whether it is the first or second PA flip, the maximal post-flip polarized flux is highest when the LOS is near the jet edge. Thus, the PA flip is most easily observed near the jet edge. In addition, the post-flip polarized flux of the first PA flip is higher in a larger parameter range than that of the second flip. By comparing these two figures, whether we change $\theta_j$ or $\Gamma$, the parameter conditions of $q\gtrsim 0.5$ (except for $q\simeq 1$) and $y_j\gtrsim 4$ remain unchanged for the PA
flip, but there are some differences in the distribution of maximal post-flip polarized flux. The major difference is that changing $\theta_j$ results in higher polarized flux under the same parameters of $q$ and $y_j$ when $\theta_j \ll 0.1$. We note that this does not mean that the smaller the $\theta_j$, the easier it is to observe the PA flip, as the smaller the jet opening angle, the smaller the corresponding observable viewing angle. However, the magnitude of the highest normalized polarized flux can only reach $10^{-4}$ whether it is changing $\theta_j$ or changing $\Gamma$. Furthermore, under most parameter values of $\Gamma$ and $\theta_j$, the highest post-flip maximal polarized flux is distributed within a narrow viewing angle range near the jet edge. Hence, we can take the typical parameters of $\theta_j=0.1$ and $\Gamma=300$ to estimate the observed rate of GRBs with PA flip.

We take the typical parameters of $\theta_j=0.1$ and $\Gamma=300$ to calculate the normalized maximal post-flip polarized flux with different $q$. The results are shown in the upper panel of Fig. \ref{FigF-t-q}. We assume a cutoff at $10^{-4}$ of the normalized polarized flux, above which it could be observed by the polarimeter.  For the first PA flip, the range of $q$ values that can observe normalized polarized flux above $10^{-4}$ is $\sim (0.9-1)$. Hence, we can roughly estimate the observed rate of GRBs with PA flip as (at least one flip)

\begin{equation}
\label{14}
R_{ch} \sim \frac{\Omega_{ch1}}{\Omega_{tot}} \sim \frac{\cos{(0.9 \theta_j)}- \cos{(\theta_j)}}{1-\cos{(1.1 \theta_j)}}
,\end{equation}
where 
\begin{equation}
\Omega_{ch} \simeq \int_{0.8\theta_j}^{1.1\theta_j} \sin{\theta} d\theta \int_{0}^{2\pi} d\phi =2\pi [\cos{(0.9 \theta_j)}- \cos{(1.1\theta_j)}]
.\end{equation}

For $\theta_j \sim 0.1$, we obtain $R_{ch} \sim \frac{\cos{(0.9 \times 0.1)}- \cos{ 0.1}}{1-\cos{(1.1\times 0.1)}}\sim 16 \% $.  For the second PA flip, the range of $q$ values that can observe normalized polarized flux above $10^{-4}$ is $\sim (0.95-1)$ (see Fig. \ref{FigF-t-q}); thus, we can give the observed rate of GRBs with double 90-degree PA flips as
\begin{equation}
R_{ch2} \sim \frac{\Omega_{ch2}}{\Omega_{tot}} \sim \frac{\cos{(0.95 \theta_j)}- \cos{(\theta_j)}}{1-\cos{(1.1 \theta_j)}}
,\end{equation}
where 
\begin{equation}
\Omega_{ch2} \simeq \int_{0.9\theta_j}^{\theta_j} \sin{\theta} d\theta \int_{0}^{2\pi} d\phi =2\pi [\cos{(0.95 \theta_j)}- \cos{\theta_j}]
.\end{equation}
Taking $\theta_j \sim 0.1$, we have $R_{ch2} \sim \frac{\cos{(0.95 \times 0.1)}- \cos{ (0.1)}}{1-\cos{(1.1\times 0.1)}}\sim 8 \% $.
Furthermore, we can estimate the observed rate of GRBs with a single PA flip as $R_{ch1}\simeq R_{ch} - R_{ch2} \sim 16\% -8\% \sim 8\%$. It is noted that there is no PA change when $q\simeq 1$. Thus, the observed rate would in reality be a little lower than the estimation above. Moreover, we should note that the observed rate in the $\gamma$-ray band we estimated above is also applicable for the case in the X-ray band, but it is not accurate enough to be applied for the case in the optical band since the flux in the optical band declines much slower than that in the X-ray or the $\gamma$-ray band (see Fig.\ref{flux_pola_t}). However, the observed rate mainly depends on the temporal resolution of polarimeters. This is because the higher the temporal resolution of the polarimeters, the wider the range of viewing angles that can be obtained for a 90-degree PA flip.

The observed rate we estimated above is just for the single pulse GRBs, but the multiple pulses are common in GRB light curves. For the multiple-pulse GRBs, due to the overlap between pulses, the high-latitude radiation of earlier pulses would be covered by the low-latitude radiation of subsequent pulses. As discussed in Section \ref{subsec:3.1}, the radiation mainly comes from high latitudes when the PA changes by 90 degrees for a single pulse. Therefore, the PA flip should occur within the tail of the last pulse for multiple pulse bursts.  Thus, the PA changes in a multiple-pulse burst are mainly determined by the last pulse. This deduction is also supported by \citet{Gill+Granot+2021}. Hence, the observed rate of the multiple-pulse GRBs with PA change of 90 degrees should approach that of the single-pulse bursts. 

We also calculate the time at PA flip ($t_{obs,ch} / t_{peak}$, normalized to the peak time) with different $q$. These results are shown in the bottom panel of Fig. \ref{FigF-t-q}. Considering that it is easiest to observe GRBs with PA flip when the LOS near the edge of the jet, the first and second PA flips in a GRB pulse are most likely to occur at the time of $t_{\rm{obs}} \sim [2-3] t_{\rm{peak}}$ and $\sim [3-4] t_{\rm{peak}}$ (see Fig. \ref{FigF-t-q}).

\section{Summary and discussion}
\label{sec:5}

\begin{figure*}
  \centering
  \includegraphics[width=\textwidth]{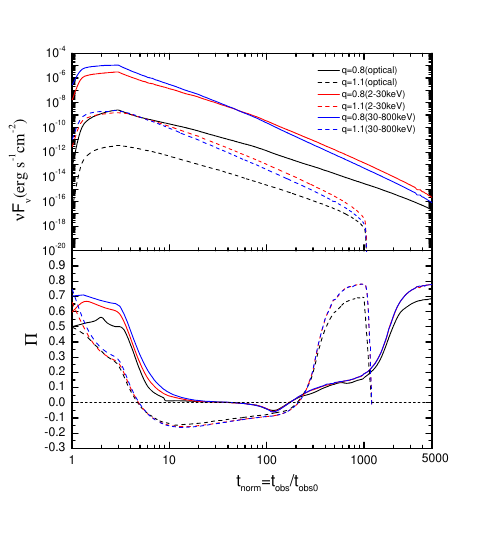}
     \caption{Comparisons of 
     fluxes (upper panel) and instantaneous PDs (bottom panel) of different energy ranges for $q=0.8$ and $q=1.1$.  
                          }
        \label{flux_pola_t}
  \end{figure*}

\begin{figure*}
   \centering
   \includegraphics[width=\textwidth]{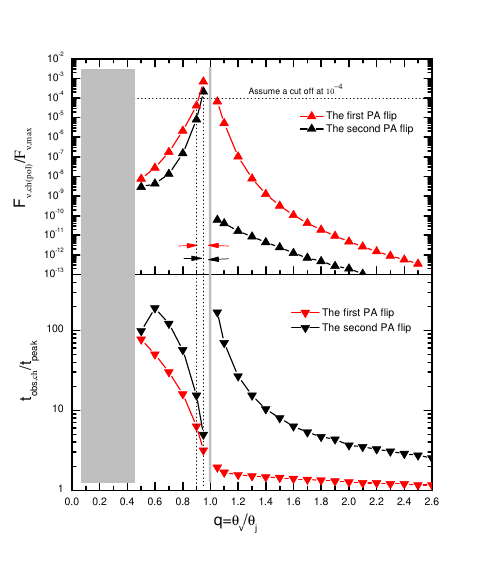}
      \caption{This plot is for estimating the observed rate of GRBs with 90-degree PA flips and the time nodes of these flips. Upper panel shows the normalized maximal post-flip polarized flux with different $q$ by taking typical parameters of $\Gamma=300$ and $\theta_j=0.1$. The bottom panel shows the normalized time at PA flip ($t_{obs,ch} / t_{peak}$) with different $q$. 
      The gray regions represent that there are no PA flips in these areas. 
                                }
         \label{FigF-t-q}
   \end{figure*}

In this paper, we examine the synchrotron polarization evolution during GRB prompt emission by taking the following assumptions. These include the MF being a perfectly ordered large-scale toroidal field that decreases as the radius increases with an assumed power-law slope and the jet being homogeneous. The model we considered can produce spectra similar to the observed GRB spectra \citep{Uhm+Zhang+2014}. We identify the parameter conditions under which the PA encounters 90-degree changes and estimate the observation rate of the bursts exhibiting the PA variations. If the polarization evolution mode we derived is verified to be similar to the future observations, it would support the current model. Our main results are summarized as follows.
\begin{enumerate}

\item The 90-degree PA flip in the toroidal MF in GRB prompt emission happens for the following two reasons. One is the low level of symmetry of the observed area about the direction of the LOS for $q>0.5$ compared to the case of $q<0.5$. The other one is the radiation disappearance in local regions in the viewing field, which could change the dominant direction of the polarization vectors.
 \item The PA rotation is related to the factors of the viewing angle, jet opening angle, and jet Lorentz factor. The parameter conditions of the 90-degree PA flip are $q \gtrsim 0.5$ (except for $q \simeq 1$) and $y_j =(\Gamma \theta_j)^2 \gtrsim 4$.
 \item By taking an assumption of a cutoff at $10^{-4}$ of the normalized polarized flux above which it could be observed by the polarimeter, the observed rate of GRBs with this PA flip in X-ray or $\gamma$-ray band is estimated as $R_{ch} \sim 16\%$, where the observed rates of single and double flips each account for $\sim 8\%$. We note that these results apply to both single-pulse and multiple-pulse GRBs.
 
\item When the LOS approaches the jet edge, it offers optimal conditions for detecting the 90-degree PA flip, primarily due to the high normalized polarized flux of $F_{\rm{\nu, ch(pol)}}/ F_{\rm{\nu,max}}$ in this case.

\item The first and second PA flips in a GRB pulse are most likely to occur at the time of $t_{obs} \sim [2-3] t_{peak}$ and $\sim [3-4] t_{peak}$ according to Fig. \ref{FigF-t-q}. It is also noted that the PA flip would not happen before the peak time $t_{\rm{peak}}$.

\item Compared to the X-ray and $\gamma$-ray bands, the second 90-degree PA changes should be more easily observed in the optical band because the flux in the optical band declines much slower than that in the X-ray or $\gamma$-ray band.
\\ \hspace*{\fill} \\
We should note that the observed rate of GRBs with a 90-degree PA flip is a rough estimation in this paper. It is based on our assumption that setting a cutoff at $10^{-4}$ of the normalized polarized flux above which it could be observed by the polarimeter. In fact, it is not easy to measure this PA flip. This is attributed not only to the low post-flip polarized flux, but also to the difficulties from the background (low signal-to-noise level at such low fluxes). Considering these factors, it will require some favorable conditions to observe this effect, and the observed rate of this PA flip could be much lower than $16\%$. If the rate observed in the future is consistent with this estimation, it would support the toroidal MF model in GRB prompt emission. Conversely, if the actual observation rate is much higher than this estimation in the future, then there could be other mechanisms that are responses to the 90-degree PA flip. This needs to be verified by the observations of high-precision polarimeters in the future, such as POLAR-2 and LEAP. It also should be noted that detecting the second 90-degree change in PA is typically challenging due to the relatively low flux at the time of this change. We expect that the optical or X-ray polarimeters (such as RINGO3 and POLAR-2) will observe GRBs twice exhibiting PA changes of 90 degrees in the future. This is because the flux in the lower energy bands declines much shallower compared to that in the higher energy bands (see Fig. \ref{flux_pola_t}).
\\ \hspace*{\fill} \\
If we observe GRBs with the PA changing by 90 degrees, we may expect that the LOS is likely near the edge of the jet. As a result, the first and second PA changes in a GRB pulse are most likely to happen at the time of $t_{obs} \sim [2-3] t_{peak}$ and $\sim [3-4] t_{peak}$ according to Fig. \ref{FigF-t-q}. Following the time-resolved analysis of Burgess19, the initial PA of GRB 170114A is $\sim 70$ degrees, and the PA at the peak time ($t_{\rm{peak}} \sim 1.8$s) is $\sim 60$ degrees. Thus, there are no 90-degree PA changes before the peak time, and the changes in PA are small. The PA at the time of $t_{obs}$ $\sim 3.6$s $\sim 2 t_{peak}$ is $\sim 150$ degrees. Thus, the PA changes by $\sim 90$ degrees compared with the initial PA. This is the first PA change by 90 degrees. Subsequently, the PA at the time of $t_{obs}$ $\sim 5.4$s $\sim 3 t_{peak}$ is $\sim 50$ degrees. The PA changes by $\sim 90$ degrees compared with the PA at the time of $t_{obs} \sim 2 t_{peak}$. This is the second 90-degree PA change. Hence, the time-resolved results of PAs from our calculation are roughly consistent with the time-resolved PAs shown in GRB 170114A in Burgess19. The theoretic results need to be further verified by more observational data in the future. 

\end{enumerate}

\begin{acknowledgements}
We thank Jianchao Sun for the useful discussion. This work is supported by the Guangxi Natural Science Foundation (GUIKEAD22035945), the Guangxi Natural Science Foundation (2024GXNSFBA010350), the Scientific Research Project of Guangxi Minzu University (2021KJQD03), the National Key R\&D Program of China (2023YFE0101200), the Guangxi Natural Science Foundation (2019GXNSFFA245008), and the National Natural Science Foundation of China (Nos. U1831135, 12393813, 12393811).
J.M is supported by the Yunnan
Revitalization Talent Support Program (YunLing Scholar Project).
\end{acknowledgements}

\bibliographystyle{a_a}
\bibliography{a_a}

\begin{thebibliography}{50}
\expandafter\ifx\csname natexlab\endcsname\relax\def\natexlab#1{#1}\fi

\bibitem[{{Arnold} {et~al.}(2012){Arnold}, {Steele}, {Bates}, {Mottram}, \&
  {Smith}}]{Arnold+etal+2012}
{Arnold}, D.~M., {Steele}, I.~A., {Bates}, S.~D., {Mottram}, C.~J., \& {Smith},
  R.~J. 2012, in Society of Photo-Optical Instrumentation Engineers (SPIE)
  Conference Series, Vol. 8446, Ground-based and Airborne Instrumentation for
  Astronomy IV, ed. I.~S. {McLean}, S.~K. {Ramsay}, \& H.~{Takami}, 84462J

\bibitem[{{Burgess} {et~al.}(2019){Burgess}, {Kole}, {Berlato}, {Greiner},
  {Vianello}, {Produit}, {Li}, \& {Sun}}]{Burgess+etal+2019}
{Burgess}, J.~M., {Kole}, M., {Berlato}, F., {et~al.} 2019, A\&A, 627, A105

\bibitem[{{Chang} \& {Lin}(2014)}]{Chang+Lin+2014}
{Chang}, Z. \& {Lin}, H.-N. 2014, MNRAS, 445, 4105

\bibitem[{{Cheng} {et~al.}(2020){Cheng}, {Zhao}, \& {Bai}}]{Cheng+etal+2020}
{Cheng}, K.~F., {Zhao}, X.~H., \& {Bai}, J.~M. 2020, \mnras, 498, 3492

\bibitem[{{Coburn} \& {Boggs}(2003)}]{Coburn+Boggs+2003}
{Coburn}, W. \& {Boggs}, S.~E. 2003, \nat, 423, 415

\bibitem[{{Deng} {et~al.}(2016){Deng}, {Zhang}, {Zhang}, \&
  {Li}}]{Deng+etal+2016}
{Deng}, W., {Zhang}, H., {Zhang}, B., \& {Li}, H. 2016, \apjl, 821, L12

\bibitem[{{Fan} {et~al.}(2023){Fan}, {Liu}, {Feng}, {Xie}, {Wang}, {Chen},
  {Feng}, {Xie}, \& {Liang}}]{Fan+etal+2023}
{Fan}, Z., {Liu}, H., {Feng}, H., {et~al.} 2023, IEEE Transactions on Nuclear
  Science, 70, 1507

\bibitem[{{Geng} {et~al.}(2018){Geng}, {Huang}, {Wu}, {Zhang}, \&
  {Zong}}]{Geng+etal+2018}
{Geng}, J.-J., {Huang}, Y.-F., {Wu}, X.-F., {Zhang}, B., \& {Zong}, H.-S. 2018,
  ApJS, 234, 3

\bibitem[{{Gill} \& {Granot}(2021)}]{Gill+Granot+2021}
{Gill}, R. \& {Granot}, J. 2021, \mnras, 504, 1939

\bibitem[{{Gill} \& {Granot}(2024)}]{Gill+Granot+2024}
{Gill}, R. \& {Granot}, J. 2024, \mnras, 527, 12178

\bibitem[{{Gill} {et~al.}(2020){Gill}, {Granot}, \& {Kumar}}]{Gill+etal+2020}
{Gill}, R., {Granot}, J., \& {Kumar}, P. 2020, MNRAS, 491, 3343

\bibitem[{{Gill} {et~al.}(2021){Gill}, {Kole}, \& {Granot}}]{Gill+etal+2021}
{Gill}, R., {Kole}, M., \& {Granot}, J. 2021, Galaxies, 9, 82

\bibitem[{{Granot}(2003)}]{Granot+2003}
{Granot}, J. 2003, \apjl, 596, L17

\bibitem[{{Granot} \& {K{\"o}nigl}(2003)}]{Granot+konigl+2003}
{Granot}, J. \& {K{\"o}nigl}, A. 2003, \apjl, 594, L83

\bibitem[{{Granot} \& {Taylor}(2005)}]{Granot+Taylor+2005}
{Granot}, J. \& {Taylor}, G.~B. 2005, \apj, 625, 263

\bibitem[{{Gruzinov} \& {Waxman}(1999)}]{Gruzinov+Waxman+1999}
{Gruzinov}, A. \& {Waxman}, E. 1999, ApJ, 511, 852

\bibitem[{{Kole}(2019)}]{Kole+etal+2019}
{Kole}, M. 2019, in International Cosmic Ray Conference, Vol.~36, 36th
  International Cosmic Ray Conference (ICRC2019), 572

\bibitem[{{Kole} {et~al.}(2020){Kole}, {De Angelis}, {Berlato}, {Burgess},
  {Gauvin}, {Greiner}, {Hajdas}, {Li}, {Li}, {Pollo}, {Produit}, {Rybka},
  {Song}, {Sun}, {Szabelski}, {Tymieniecka}, {Wang}, {Wu}, {Wu}, {Xiong},
  {Zhang}, \& {Zhang}}]{Kole+etal+2020}
{Kole}, M., {De Angelis}, N., {Berlato}, F., {et~al.} 2020, A\&A, 644, A124

\bibitem[{{Lan} \& {Dai}(2020)}]{Lan+etal+2020}
{Lan}, M.-X. \& {Dai}, Z.-G. 2020, \apj, 892, 141

\bibitem[{{Lan} {et~al.}(2019){Lan}, {Geng}, {Wu}, \& {Dai}}]{Lan+etal+2019}
{Lan}, M.-X., {Geng}, J.-J., {Wu}, X.-F., \& {Dai}, Z.-G. 2019, \apj, 870, 96

\bibitem[{{Lan} {et~al.}(2021{\natexlab{a}}){Lan}, {Wang}, {Xu}, {Liu}, \&
  {Wu}}]{Lan+etal+2021(2)}
{Lan}, M.-X., {Wang}, H.-B., {Xu}, S., {Liu}, S., \& {Wu}, X.-F.
  2021{\natexlab{a}}, \apj, 909, 184

\bibitem[{{Lan} {et~al.}(2021{\natexlab{b}}){Lan}, {Wu}, \&
  {Dai}}]{Lan+etal+2021}
{Lan}, M.-X., {Wu}, X.-F., \& {Dai}, Z.-G. 2021{\natexlab{b}}, Research in
  Astronomy and Astrophysics, 21, 055

\bibitem[{{Lazzati}(2006)}]{Lazzati+2006}
{Lazzati}, D. 2006, New Journal of Physics, 8, 131

\bibitem[{{Longair}(2011)}]{Longair+2011}
{Longair}, M.~S. 2011, {High Energy Astrophysics}

\bibitem[{{Mao} \& {Wang}(2011)}]{mao11}
{Mao}, J. \& {Wang}, J. 2011, ApJ, 731, 26

\bibitem[{{Mao} \& {Wang}(2013)}]{Mao+Wang+2013}
{Mao}, J. \& {Wang}, J. 2013, \apj, 776, 17

\bibitem[{{Mao} \& {Wang}(2018)}]{Mao+Wang+2018}
{Mao}, J. \& {Wang}, J. 2018, ApJ, 854, 51

\bibitem[{{McConnell} {et~al.}(2020){McConnell}, {Baring}, {Bloser}, {Briggs},
  {Dwyer}, {Foucart}, {Gaskin}, {Goldstein}, {Grove}, {Gunji}, {Hartmann},
  {Hill-Kittle}, {Hui}, {Kippen}, {Kislat}, {Kocevski}, {Kole}, {Krizmanic},
  {Legere}, {Littenberg}, {Lyutikov}, {Mattingly}, {McBreen}, {Meegan},
  {Paciesas}, {Pearce}, {Preece}, {Prescod-Weinstein}, {Produit}, {Ryan},
  {Ryde}, {Sturner}, {Toma}, {Vestrand}, {Wilson-Hodge}, {Zhang}, \&
  {Zhang}}]{McConnel+etal+2020}
{McConnell}, M., {Baring}, M., {Bloser}, P., {et~al.} 2020, in American
  Astronomical Society Meeting Abstracts, Vol. 235, American Astronomical
  Society Meeting Abstracts \#235, 373.08

\bibitem[{{Medvedev} \& {Loeb}(1999)}]{Medvedev+Loeb+1999}
{Medvedev}, M.~V. \& {Loeb}, A. 1999, ApJ, 526, 697

\bibitem[{{Nakar} {et~al.}(2003){Nakar}, {Piran}, \&
  {Waxman}}]{Nakar+etal+2003}
{Nakar}, E., {Piran}, T., \& {Waxman}, E. 2003, JCAP, 2003, 005

\bibitem[{{Parsotan} {et~al.}(2020){Parsotan}, {L{\'o}pez-C{\'a}mara}, \&
  {Lazzati}}]{Parsotan+etal+2020}
{Parsotan}, T., {L{\'o}pez-C{\'a}mara}, D., \& {Lazzati}, D. 2020, ApJ, 896,
  139

\bibitem[{{Rossi} {et~al.}(2004){Rossi}, {Lazzati}, {Salmonson}, \&
  {Ghisellini}}]{Rossi+etal+2004}
{Rossi}, E.~M., {Lazzati}, D., {Salmonson}, J.~D., \& {Ghisellini}, G. 2004,
  \mnras, 354, 86

\bibitem[{{Rutledge} \& {Fox}(2004)}]{Rutledge+Fox+2004}
{Rutledge}, R.~E. \& {Fox}, D.~B. 2004, \mnras, 350, 1288

\bibitem[{{Rybicki} \& {Lightman}(1979)}]{Rybicki+Lightman+1979}
{Rybicki}, G.~B. \& {Lightman}, A.~P. 1979, {Radiative processes in
  astrophysics}

\bibitem[{{Sharma} {et~al.}(2019){Sharma}, {Iyyani}, {Bhattacharya},
  {Chattopadhyay}, {Rao}, {Aarthy}, {Vadawale}, {Mithun}, {Bhalerao}, {Ryde},
  \& {Pe'er}}]{Sharma+etal+2019}
{Sharma}, V., {Iyyani}, S., {Bhattacharya}, D., {et~al.} 2019, ApJL, 882, L10

\bibitem[{{Singh} {et~al.}(2014){Singh}, {Tandon}, {Agrawal}, {Antia},
  {Manchanda}, {Yadav}, {Seetha}, {Ramadevi}, {Rao}, {Bhattacharya}, {Paul},
  {Sreekumar}, {Bhattacharyya}, {Stewart}, {Hutchings}, {Annapurni}, {Ghosh},
  {Murthy}, {Pati}, {Rao}, {Stalin}, {Girish}, {Sankarasubramanian},
  {Vadawale}, {Bhalerao}, {Dewangan}, {Dedhia}, {Hingar}, {Katoch}, {Kothare},
  {Mirza}, {Mukerjee}, {Shah}, {Shah}, {Mohan}, {Sangal}, {Nagabhusana},
  {Sriram}, {Malkar}, {Sreekumar}, {Abbey}, {Hansford}, {Beardmore}, {Sharma},
  {Murthy}, {Kulkarni}, {Meena}, {Babu}, \& {Postma}}]{Singh+etal+2014}
{Singh}, K.~P., {Tandon}, S.~N., {Agrawal}, P.~C., {et~al.} 2014, in Society of
  Photo-Optical Instrumentation Engineers (SPIE) Conference Series, Vol. 9144,
  Space Telescopes and Instrumentation 2014: Ultraviolet to Gamma Ray, ed.
  T.~{Takahashi}, J.-W.~A. {den Herder}, \& M.~{Bautz}, 91441S

\bibitem[{{Spruit} {et~al.}(2001){Spruit}, {Daigne}, \&
  {Drenkhahn}}]{Spruit+2001}
{Spruit}, H.~C., {Daigne}, F., \& {Drenkhahn}, G. 2001, A \& A, 369, 694

\bibitem[{{Tavecchio} {et~al.}(2003){Tavecchio}, {Ghisellini}, \&
  {Celotti}}]{Tavecchio+etal+2003}
{Tavecchio}, F., {Ghisellini}, G., \& {Celotti}, A. 2003, A \& A, 403, 83

\bibitem[{{Toma} {et~al.}(2009){Toma}, {Sakamoto}, {Zhang}, {Hill},
  {McConnell}, {Bloser}, {Yamazaki}, {Ioka}, \& {Nakamura}}]{Toma+etal+2009}
{Toma}, K., {Sakamoto}, T., {Zhang}, B., {et~al.} 2009, ApJ, 698, 1042

\bibitem[{{Uhm} \& {Zhang}(2014)}]{Uhm+Zhang+2014}
{Uhm}, Z.~L. \& {Zhang}, B. 2014, Nature Physics, 10, 351

\bibitem[{{Uhm} {et~al.}(2012){Uhm}, {Zhang}, {Hasco{\"e}t}, {Daigne},
  {Mochkovitch}, \& {Park}}]{Uhm+etal+2012}
{Uhm}, Z.~L., {Zhang}, B., {Hasco{\"e}t}, R., {et~al.} 2012, ApJ, 761, 147

\bibitem[{{Wang} \& {Lan}(2023{\natexlab{a}})}]{Wang+Lan+2023}
{Wang}, H.-B. \& {Lan}, M.-X. 2023{\natexlab{a}}, \apj, 946, 12

\bibitem[{{Wang} \& {Lan}(2023{\natexlab{b}})}]{Wang+Lan+2023-2}
{Wang}, H.-B. \& {Lan}, M.-X. 2023{\natexlab{b}}, arXiv e-prints,
  arXiv:2306.16618

\bibitem[{{Wigger} {et~al.}(2004){Wigger}, {Hajdas}, {Arzner}, {G{\"u}del}, \&
  {Zehnder}}]{Wigger+etal+2004}
{Wigger}, C., {Hajdas}, W., {Arzner}, K., {G{\"u}del}, M., \& {Zehnder}, A.
  2004, \apj, 613, 1088

\bibitem[{{Xie} {et~al.}(2022){Xie}, {Di Marco}, {La Monaca}, {Liu}, {Muleri},
  {Bucciantini}, {Romani}, {Costa}, {Rankin}, {Soffitta}, {Bachetti}, {Di
  Lalla}, {Fabiani}, {Ferrazzoli}, {Gunji}, {Latronico}, {Negro}, {Omodei},
  {Pilia}, {Trois}, {Watanabe}, {Agudo}, {Antonelli}, {Baldini}, {Baumgartner},
  {Bellazzini}, {Bianchi}, {Bongiorno}, {Bonino}, {Brez}, {Capitanio},
  {Castellano}, {Cavazzuti}, {Ciprini}, {De Rosa}, {Del Monte}, {Di Gesu},
  {Donnarumma}, {Doroshenko}, {Dov{\v{c}}iak}, {Ehlert}, {Enoto},
  {Evangelista}, {Garcia}, {Hayashida}, {Heyl}, {Iwakiri}, {Jorstad}, {Karas},
  {Kitaguchi}, {Kolodziejczak}, {Krawczynski}, {Liodakis}, {Maldera},
  {Manfreda}, {Marin}, {Marinucci}, {Marscher}, {Marshall}, {Massaro}, {Matt},
  {Mitsuishi}, {Mizuno}, {Ng}, {O'Dell}, {Oppedisano}, {Papitto}, {Pavlov},
  {Peirson}, {Perri}, {Pesce-Rollins}, {Petrucci}, {Possenti}, {Poutanen},
  {Puccetti}, {Ramsey}, {Ratheesh}, {Sgr{\'o}}, {Slane}, {Spandre}, {Tamagawa},
  {Tavecchio}, {Taverna}, {Tawara}, {Tennant}, {Thomas}, {Tombesi},
  {Tsygankov}, {Turolla}, {Vink}, {Weisskopf}, {Wu}, \& {Zane}}]{Xie+etal+2022}
{Xie}, F., {Di Marco}, A., {La Monaca}, F., {et~al.} 2022, \nat, 612, 658

\bibitem[{{Yonetoku} {et~al.}(2012){Yonetoku}, {Murakami}, {Gunji}, {Mihara},
  {Toma}, {Morihara}, {Takahashi}, {Wakashima}, {Yonemochi}, {Sakashita},
  {Toukairin}, {Fujimoto}, \& {Kodama}}]{Yonetoku+etal+2012}
{Yonetoku}, D., {Murakami}, T., {Gunji}, S., {et~al.} 2012, ApJL, 758, L1

\bibitem[{{Yonetoku} {et~al.}(2011){Yonetoku}, {Murakami}, {Gunji}, {Mihara},
  {Toma}, {Sakashita}, {Morihara}, {Takahashi}, {Toukairin}, {Fujimoto},
  {Kodama}, {Kubo}, \& {IKAROS Demonstration Team}}]{Yonetoku+etal+2011}
{Yonetoku}, D., {Murakami}, T., {Gunji}, S., {et~al.} 2011, ApJL, 743, L30

\bibitem[{{Zhang} \& {Yan}(2011)}]{Zhang+Yan+2011}
{Zhang}, B. \& {Yan}, H. 2011, \apj, 726, 90

\bibitem[{{Zhang} {et~al.}(2019){Zhang}, {Kole}, {Bao}, {Batsch}, {Bernasconi},
  {Cadoux}, {Chai}, {Dai}, {Dong}, {Gauvin}, {Hajdas}, {Lan}, {Li}, {Li}, {Li},
  {Liu}, {Liu}, {Marcinkowski}, {Produit}, {Orsi}, {Pohl}, {Rybka}, {Shi},
  {Song}, {Sun}, {Szabelski}, {Tymieniecka}, {Wang}, {Wang}, {Wen}, {Wu}, {Wu},
  {Wu}, {Xiao}, {Xiong}, {Zhang}, {Zhang}, {Zhang}, {Zhang}, \&
  {Zwolinska}}]{Zhang+etal+2019}
{Zhang}, S.-N., {Kole}, M., {Bao}, T.-W., {et~al.} 2019, Nature Astronomy, 3,
  258

\bibitem[{{Zhao} {et~al.}(2014){Zhao}, {Li}, {Liu}, {Zhang}, {Bai}, \&
  {M{\'e}sz{\'a}ros}}]{Zhao+etal+2014}
{Zhao}, X., {Li}, Z., {Liu}, X., {et~al.} 2014, ApJ, 780, 12

\end{thebibliography}


\end{document}